\documentclass[epsfig,usenatbib]{mnras}
\usepackage{epsfig}
\usepackage{natbib}
\usepackage{amsmath}

\def\gtsima{$\; \buildrel > \over \sim \;$}
\def\ltsima{$\; \buildrel < \over \sim \;$}
\def\gtrsim{\lower.5ex\hbox{\gtsima}}
\def\lesssim{\lower.5ex\hbox{\ltsima}}

\begin{document}
%

\title[Black hole mergers in the Illustris]{The cosmic merger rate of stellar black hole binaries from the Illustris simulation}
\author[Mapelli et al.]
{Michela Mapelli$^{1,2}$, Nicola Giacobbo$^{1,3}$, Emanuele Ripamonti$^3$, Mario Spera$^{1,2,4}$ 
\\
$^1$INAF-Osservatorio Astronomico di Padova, Vicolo dell'Osservatorio 5, I--35122, Padova, Italy, {\tt michela.mapelli@oapd.inaf.it}\\
$^2$INFN, Milano Bicocca, Piazza della Scienza 3, I-20126 Milano, Italy\\
$^3$Physics and Astronomy Department Galileo Galilei, University of Padova, Vicolo dell'Osservatorio 3, I--35122, Padova, Italy\\
$^4$Physics Department Giuseppe Occhialini, University of Milano Bicocca, Piazza della Scienza 3, I-20126 Milano, Italy\\
}
\maketitle \vspace {7cm }
\bibliographystyle{mnras}
 
\begin{abstract}
The cosmic merger rate density of black hole binaries (BHBs) can give us an essential clue to constraining the formation channels of BHBs, in light of current and forthcoming  gravitational wave detections. Following a Monte Carlo approach, we couple new population-synthesis models of BHBs with the Illustris cosmological simulation, to study the cosmic history of BHB mergers. We explore six population-synthesis models, varying the prescriptions for supernovae, common envelope, and natal kicks. In most considered models, the cosmic BHB merger rate follows the same trend as the cosmic star formation rate. The normalization of the cosmic BHB merger rate strongly depends on the treatment of common envelope and on the distribution of natal kicks. We find that most BHBs merging within LIGO's instrumental horizon come from relatively metal-poor progenitors ($<0.2$ Z$_\odot$). The total masses of merging BHBs span a large range of values, from $\sim{}6$ to $\sim{}82$ M$_\odot$.  In our fiducial model, merging BHBs consistent with GW150914, GW151226 and GW170104 represent $\sim{}6$, $3$, and $12$ per cent of all BHBs merging within the LIGO horizon, respectively. The heavy systems, like GW150914, come from metal-poor progenitors ($<0.15$ Z$_\odot$). Most GW150914-like systems merging in the local Universe appear to have formed at high redshift, with a long delay time. In contrast, GW151226-like systems form and merge all the way through the cosmic history, from progenitors with a broad range of metallicities. Future detections will be crucial to put  constraints on common envelope, on natal kicks, and on the BHB mass function.  
\end{abstract}

%

\section{Introduction}
The first direct detection of gravitational waves (GWs, \citealt{abbott2016aDISCO}) opens a new perspective for the study of compact object (CO) binaries. Black hole binaries (BHBs)  have been predicted and studied  for a long time (e.g. \citealt{tutukov1973,thorne1987,schutz1989,kulkarni1993,sigurdsson1993,portegieszwart2000,colpi2003,belczynski2004}), but the  three events observed by LIGO so far  (GW150914, GW151226, and GW170104) are the first observational confirmation of their existence. 

Moreover, the two black holes (BHs) associated with GW150914  and one of the two BHs associated with GW170104 are surprisingly massive: $36.2^{+5.2}_{-3.8}\,{}{\rm M}_\odot$, $29.1^{+3.7}_{-4.4}\,{}{\rm M}_\odot$ \citep{abbott2016bASTRO,abbott2016cO1}, and $31.2^{+8.4}_{-6.0}$ M$_\odot$ \citep{abbott2017}, respectively.  If they are the remnants of massive stars, such massive BHs should have formed from relatively metal-poor ($Z\leq{}0.5$ Z$_\odot$) progenitors, which are expected to collapse directly to BHs (e.g. \citealt{mapelli2009,mapelli2010,mapelli2013,belczynski2010,spera2015}). Dynamical processes (such as exchanges or runaway collisions in dense star clusters) might also contribute to enhancing the formation of massive BHBs similar to GW150914  and GW170104 (e.g. \citealt{ziosi2014,chatterjee2016,rodriguez2015,rodriguez2016,mapelli2016}). Alternatively, GW150914  and GW170104  might be the result of primordial BHs born from gravitational collapse in the early Universe (e.g. \citealt{bird2016,carr2016,inomata2016}).

 Constraining the formation epoch and the birthplace of BHBs is one of the key points to interpret the nature of GW events associated with BHB mergers. 
This requires to model the formation and evolution of BHBs in a cosmological context. This task is currently a challenge, because of the huge dynamical range between the scale of cosmological structures (tens of Mpc) and the scale of binary evolution  ($\lesssim{}$ few AU). Moreover, since the progenitor's metallicity appears to be crucial for the BH mass \citep{spera2015}, any attempt to reconstruct the cosmic formation of BHBs should account for the local and global evolution of metallicity in the proper way.

For these reasons, only few authors attempted to put the formation of  BHBs in a cosmological frame. \cite{dominik2013} plant CO binaries 
into the cosmic history through a Monte Carlo-based algorithm. They generate a sample of galaxies 
based on a Press-Schechter like function \citep{fontana2006}, 
adopt the average metallicity evolution  described by \cite{pei1999}, and finally associate the CO binaries to a given redshift bin based on the cosmic star formation rate (SFR) evolution \citep{strolger2004}. This gives a BHB merger rate density reaching its maximum at $z\sim{}4-5$ and then slowly decreasing down to $z=0$. 

Similarly, \cite{belczynski2016}  generate isolated BHBs 
and then distribute them as a function of redshift, adopting an updated version of the cosmic SFR density and of the average metallicity evolution \citep{madaudickinson2014}. 
This approach does not account for the mass-metallicity relation observed in galaxies \citep{maiolino2008}. The resulting merger rate density peaks at $z\sim{}2$. If only GW150914-like systems are considered, the distribution of the formation times of these systems is markedly bimodal with two peaks, one $\sim{}11-12$ Gyr ago and the second one $\sim{}2-3$ Gyr ago.


In contrast, \cite{lamberts2016} 
account for the cosmological evolution through a Press-Schechter like formalism \citep{cole2008} with a redshift-dependent mass-metallicity relation \citep{ma2016}. This ensures that the metallicity of a galaxy depends on its mass, consistent with the observations \citep{maiolino2008}. 
 \cite{lamberts2016} do not recover the strongly bimodal birth-time distribution of GW150914-like systems reported by \cite{belczynski2016}. Their predicted BHB merger rate is $\sim{}850$ Gpc$^{-3}$ yr$^{-1}$, significantly larger than inferred from LIGO observations ($\sim{}9-240$ Gpc$^{-3}$ yr$^{-1}$, \citealt{abbott2016cO1}). A conceptually similar approach was followed also by \cite{dvorkin2016} and \cite{elbert2017}. 

The formalism adopted by \cite{dominik2013}, \cite{belczynski2016}, and even \cite{lamberts2016} cannot give us detailed information on the evolution of the host galaxy of a CO binary. 
Thus, \cite{oshaughnessy2017} follow a complementary approach: they start from a cosmological simulation and pick up four test galaxies,  
which they re-simulate at high resolution,  by doing a ``zoom-in''. Then, they add BHBs to the location of star forming particles in the simulation. 
They find a significantly higher merger rate per unit mass in dwarf galaxies than in Milky-Way-like galaxies.

 Recently, \cite{schneider2017} characterize the formation and coalescence
sites of GW events, by coupling the metallicity-dependent binary population synthesis code \textsc{SeBa} \citep{portegieszwart1996,mapelli2013} with a $(4\,{}{\rm Mpc})^3$ simulation performed with the \textsc{GAMESH}  pipeline \citep{graziani2015, graziani2017}.   \textsc{GAMESH} interfaces an N-body simulation with a semi-analytic model for galaxy formation, and a radiative-transfer code. With this approach, \cite{schneider2017} find that the observed GW events occur most likely in star forming galaxies with stellar mass $> 10^{10}$ M$_\odot$.  





In this paper, we follow a new approach, complementary to  previous work: we draw the cosmic history of the BHB merger rate by coupling up-to-date population synthesis simulations of BHBs  with the public Illustris-1 cosmological simulation \citep{vogelsberger2014a}. The Illustris-1 is the highest resolution hydrodynamical simulation run in the frame of the Illustris project \citep{vogelsberger2014b}.  In the following, we refer to the Illustris-1 simply as Illustris. The Illustris box (length $=106.5$ Mpc comoving) is considerably larger than the one adopted by \cite{schneider2017}, ensuring that we are considering a less biased portion of the Universe, even if with lower resolution. 

We plant our BHBs in the cosmological simulation through a Monte Carlo approach, based on the metallicity of star particles. Our BHBs were  generated by evolving isolated stellar binaries with a new version of the \textsc{BSE} code \citep{hurley2002}, which  includes  up-to-date recipes for stellar evolution and stellar winds 
\citep{vink2011,chen2015}. Moreover, we include the effect of pulsational pair-instability and pair-instability supernovae \citep{woosley2017} which were neglected in most previous studies. This approach allows us to follow the merger history of BHBs, accounting for the evolution of their environment. 

\section{Methods}\label{sec:methods}
To reconstruct the cosmic history of BHB mergers, we couple the Illustris simulation 
 with a large set of population-synthesis simulations of isolated binaries. The main ingredients of our model are the following.
\subsection{The BHBs}
We simulate the evolution of isolated stellar binaries through an updated version of the \textsc{BSE} code \citep{hurley2000,hurley2002}. The changes with respect to the original version of \textsc{BSE} are described in a companion paper by Giacobbo et al. (in prep.). Here we summarize the most important  prescriptions. 
Stellar winds have been updated based on the equations described in  \cite{belczynski2010}. Namely, a treatment of stellar winds following \cite{vink2001} and \cite{vink2005} is included for O-type and Wolf-Rayet stars, respectively. In this model, mass loss by stellar winds depends on metallicity, both in the main sequence (MS) and in later evolutionary stages.

With respect to  \cite{belczynski2010}, there is one crucial update: we take into account the dependence of the mass loss $\dot{M}$ on the electron-scattering Eddington ratio $\Gamma$ \citep{graefener2008,vink2011,vink2016}. Following \cite{chen2015},  the mass loss scales as $\dot{M}\propto{}Z^\alpha{}$ (where $Z$ is the star metallicity) with $\alpha{}=0.85$ if the electron-scattering Eddington ratio of a star is $\Gamma{}<2/3$, and $\alpha{}=2.45-2.4\,{}\Gamma{}$ if $\Gamma{}\ge{}2/3$. This ensures that the dependence of  mass loss on  metallicity almost vanishes  if the star is radiation-pressure dominated. With this relatively small change, we obtain a mass spectrum of BHs similar to the one published by \cite{spera2015} and based on the \textsc{PARSEC} stellar evolution tracks \citep{bressan2012,tang2014,chen2015}. 

Our new version of \textsc{BSE} also includes new fitting formulas for  the core radii, as described in \cite{halltout2014}. This is a crucial ingredient for the study of BHBs, because the fate of a common envelope phase depends on the core radius. 

Furthermore, we included in \textsc{BSE} new recipes for core-collapse supernovae (SNe). In particular, we implemented both the rapid (R) and the delayed (D) models for SN explosion presented by \cite{fryer2012}. In Appendix~\ref{sec:appendix1}, we detail the prescriptions for the CO mass in the rapid and in the delayed SN model. Finally, we added a formalism to account for pair-instability and pulsational pair-instability SNe, following \cite{spera2017} (see also \citealt{belczynski2016pair,woosley2017}).

With the new code, we ran  six sets of population-synthesis simulations. The details of the  six sets are given in  Table~\ref{tab:table1}. In particular, the simulation set labelled as `R' adopts the rapid SN model, while all the others (labelled as `D') adopt the delayed model for core-collapse SNe. Pulsational pair instability and pair instability SNe are included in all runs.

For the common envelope (CE) phase, we use the same formalism as described in \cite{hurley2002}, which depends on two free parameters, $\alpha$ and $\lambda{}$. According to this formalism, $\alpha{}$ is the fraction of binding energy converted into kinetic energy of the envelope, while $\lambda{}$  describes the geometry of the envelope. In the formalism by \cite{hurley2002}, $\alpha$ and $\lambda$ always appear as their product  $\alpha\,{}\lambda$. 
In simulations D, R, DHG, and DK we use $\alpha{}=1$, $\lambda{}=0.1$. The latter choice of $\lambda{}$ is quite well motivated for massive stars   (e.g. \citealt{xu2010,loveridge2011}).  In simulation D0.02  we assume $\alpha{}=0.2$, $\lambda{}=0.1$, while in simulation D1.5 we use  $\alpha{}=3$ and $\lambda{}=0.5$. 

The treatment of Hertzsprung gap (HG) donors was found to be critical in previous studies (e.g. \citealt{dominik2012}).  A HG star lacks a steep density gradient between the core and envelope. Thus, its response to a CE should be similar to that of a MS star \citep{ivanova2004}. In the standard version of \textsc{BSE},  MS donors entering a CE phase are forced to merge with the accretor, while HG donors are allowed to survive the CE phase. In our run DHG, we adopted the default setting of \textsc{BSE} allowing HG donors to survive a CE phase. In all the other simulations we modified \textsc{BSE}, by imposing that a HG donor merges with its companion if they enter a CE phase.

Finally, the natal kick of the CO is another essential ingredient, because it can unbind a binary. There are no conclusive observational constraints on the natal kick of BHs, even if some recent studies indicate that high-velocity kicks are possible \citep{repetto2012,repetto2017,oshaughnessy2017b}. Thus, we draw the natal kicks from the Maxwellian distribution described in \cite{hobbs2005}, with dispersion $\sigma{}=265$ km s$^{-1}$. This distribution was obtained from the proper motions of 233 isolated Galactic pulsars.  

In model DK, we assume that BH kicks follow the same distribution as neutron star (NS) kicks. 
In all the other models, we scale the velocities drawn from this distribution by the amount of fallback, according to:
\begin{equation}\label{eq:eq1}
v_{\rm BH}=v_{\rm NS} (1-f_{\rm fb}),
\end{equation}
where $v_{\rm BH}$ is the natal kick for the BH, $v_{\rm NS}$ is the natal kick for a NS (drawn from the distribution proposed by \citealt{hobbs2005}), and $f_{\rm fb}$ (ranging from 0 to 1) is the amount of  fallback on the proto-NS \citep{fryer2012,spera2015}. The definition of $f_{\rm fb}$ depends on the adopted core-collapse SN prescription. In Appendix~\ref{sec:appendix1}, we detail the values of $f_{\rm fb}$ for the two considered core-collapse SN models.  

For each of the  six simulation sets described in Table~\ref{tab:table1}, we simulate  12 sub-sets with metallicity $Z=0.0002,$ 0.0004, 0.0008, 0.0012, 0.0016, 0.002, 0.004, 0.006, 0.008, 0.012, 0.016, and 0.02.   Throughout the paper, we define solar metallicity as Z$_\odot{}=0.02$. Thus, the 12 sub-sets correspond to metallicity $Z=0.01$,   0.02,   0.04,   0.06,   0.08,   0.1,   0.2,   0.3,   0.4,   0.6,   0.8, and 1.0~Z$_\odot$. In each sub-set  we simulate $10^7$ stellar binaries. Thus, each of the  six sets of simulations is composed of $1.2\times{}10^8$ massive binaries.  The mass of the primary ($m_{\rm p}$) is randomly drawn from a Kroupa initial mass function \citep{kroupa2001} ranging\footnote{The fitting formulas by \cite{hurley2000} might be inaccurate for very massive stars. To improve the treatment of massive stars, we impose that the values of the radius of single stars are consistent with \textsc{PARSEC} stellar evolution tracks \citep{chen2015} for stars with mass $>100$ $M_\odot$, as discussed in \cite{mapelli2016}.} from 5 to 150 M$_\odot$, and the mass of the secondary ($m_{\rm s}$) is sampled according to the distribution $\mathcal{F}(q)\propto{}q^{-0.1}$ (where $q=m_{\rm s}/m_{\rm p}$) in a range  $[0.1\,{}-1]\,{}m_{\rm p}$. The orbital period $P$ and the eccentricity $e$ are randomly extracted from the distribution $\mathcal{F}(P)\propto{}(\log_{10}{P})^{-0.55}$, with $0.15\leq{}\log_{10}{(P/{\rm day})}\leq{}5.5$, and $\mathcal{F}(e)\propto{}e^{-0.4}$, with $0\leq{}e<1$, as suggested by \cite{sana2012}. 

\begin{table}
\begin{center}
\caption{\label{tab:table1}
Properties of the population-synthesis simulations.}
 \leavevmode
\begin{tabular}[!h]{cccccc}
\hline
Name & SN     & $\alpha{}$ & $\lambda{}$ & HG & Kick\\
\hline
D  & Delayed & 1.0        & 0.1         & new & F12  \\
R  & Rapid  & 1.0        & 0.1       & new & F12 \\ 
DHG  & Delayed  & 1.0        & 0.1         & \textsc{BSE} & F12 \\ 
DK  & Delayed  & 1.0        & 0.1         & new & H05 \\ 
D0.02 & Delayed  & 0.2        & 0.1         & new & F12  \\
D1.5 & Delayed  & 3.0        & 0.5         & new & F12  \\
\noalign{\vspace{0.1cm}}
\hline
\end{tabular}
\begin{flushleft}
\footnotesize{Column 1: model name; column 2: SN model (delayed and rapid from \citealt{fryer2012}); column 3: value of $\alpha{}$; column 4: value of $\lambda{}$; column 5: treatment for HG stars (`\textsc{BSE}' means same treatment as in \textsc{BSE}, `new' means that we force all CE binaries with a HG donor to merge); column 6: model for the SN kick. H05 means that we use the distribution from \cite{hobbs2005}. F12 means that we rescale the natal kicks by the fallback, as described in \cite{fryer2012}. See also equation~\ref{eq:eq1} and the text for details. }
\end{flushleft}
\end{center}
\end{table}
\subsection{The Illustris}\label{sec:section2.2}
The Illustris simulation covers a comoving volume of $(106.5\,{}{\rm Mpc})^3$, and has an initial dark matter and baryonic matter mass resolution of $6.26\times{}10^6$ and $1.26\times{}10^6$ M$_\odot$, respectively \citep{vogelsberger2014a,vogelsberger2014b}. At redshift zero the softening length is $\sim{}710$ pc, while the  smallest hydrodynamical cells have a length of 48 pc.   The large size of the Illustris' box ensures that we are modelling an unbiased portion of the Universe, satisfying the cosmological principle. The main drawback is that the population of dwarf galaxies is heavily under-resolved. In Appendix~\ref{sec:appendix2}, we estimate the impact of resolution on our main results, by comparing the Illustris-1 with the lower-resolution Illustris-3 simulation. Moreover, in a companion paper \citep{schneider2017}, we follow a complementary approach: we combine our population-synthesis models with the \textsc{GAMESH} simulation, which has a box of only $(4\,{}{\rm Mpc})^3$, but much higher resolution, in order to quantify the contribution of dwarf galaxies to the BHB merger rate.

The Illustris was run with the moving mesh code \textsc{AREPO}, to solve the inviscid Euler equations \citep{springel2010}. The Illustris includes a treatment for sub-grid physics (cooling, star formation, SNe, super-massive BH formation, accretion and merger, AGN feedback, etc), as described in \cite{vogelsberger2013}. The model of sub-grid physics adopted in the Illustris is known to produce a mass-metallicity relation \citep{genel2014,genel2016} which is sensibly steeper than the observed one (see the discussion in \citealt{vogelsberger2013} and \citealt{torrey2014}). Moreover, the simulated mass-metallicity relation does not show the observed turnover at high stellar mass ($\gtrsim{}10^{10}$ M$_\odot{}$). In Appendix~\ref{sec:appendix2} we estimate that the impact of these differences between simulated and observed mass-metallicity relation on the BHB merger rate is $\sim{}20$ per cent. 

As for the cosmology, the Illustris adopts WMAP-9 results for the cosmological parameters \citep{hinshaw2013}, that is $\Omega{}_M = 0.2726$, $\Omega{}_\Lambda = 0.7274$, $\Omega{}_b = 0.0456$, and $H_0 = 100\,{}h$ km s$^{-1}$ Mpc$^{-1}$, with $h = 0.704$.

Through  the web-based interface (API) made available by the the Illustris project ({\tt http://www.illustris-project.org/}), we downloaded the stellar particles in each snapshot, including information on their formation time, initial mass, and metallicity. A total of 108 snapshots have stellar particles, from redshift $z\sim{}16$ to 0. More details on the Illustris can be found in the presentation papers \citep{vogelsberger2014a,vogelsberger2014b}, in the release paper \citep{nelson2015} and on the aforementioned website.

\subsection{Planting BHBs into a cosmological simulation}
We wrote a Monte Carlo code to associate the simulated BHBs to the Illustris, performing the following operations. 

For each of the simulation sets listed in Table~\ref{tab:table1}, we extract information on those BHBs merging within a Hubble time. Namely, we store the BH masses and the delay time $t_{\rm delay}$ between the formation of the progenitor stellar binary and the merger of the BHB.  We also store information on the total initial stellar mass $M_{\rm BSE}$ of each sub-set of simulations with the same metallicity (including binary systems which do not evolve into BHBs). 

We read each stellar particle from the Illustris only once, when it first appears in the snapshots. We store information on its initial mass $M_{\rm Ill}$, formation redshift $z_{\rm Ill}$, and metallicity $Z_{\rm Ill}$. We then find the metallicity that best matches  $Z_{\rm Ill}$ among the 12 metallicities simulated with \textsc{BSE}\footnote{If  $Z_{\rm Ill}>0.02$ ($Z_{\rm Ill}<0.0002$), we associate to the Illustris's particle a \textsc{BSE} set with  $Z_{\rm Ill}=0.02$ ($Z_{\rm Ill}=0.0002$), since  the maximum (minimum) metallicity we simulated with \textsc{BSE} is 0.02 (0.0002). This procedure is particularly arbitrary for population~III stars, whose binarity properties are not known. However, we show in Section~\ref{sec:discussion} that population~III stars do not significantly affect the rate of detectable BHB mergers.}. 

We then associate to each Illustris' particle a number $n_{\rm BHB}$ of merging BHBs, randomly extracted from the sub-set with the best-matching metallicity, based on the following algorithm:
\begin{equation}
n_{\rm BHB}=N_{\rm BSE}\,{}\frac{M_{\rm Ill}}{M_{\rm BSE}}\,{}f_{\rm corr}\,{}f_{\rm bin},
\end{equation}
where $M_{\rm Ill}$ is the initial stellar mass of the  Illustris' particle and $M_{\rm BSE}$ is the initial stellar mass in the \textsc{BSE} sub-set with the selected metallicity.  In our calculations, $M_{\rm Ill}<M_{\rm BSE}$.  $N_{\rm BSE}$ is the number of merging BHBs within the simulated sub-set of initial stellar mass $M_{\rm BSE}$. $f_{\rm corr}= 0.285$ is a correction factor, accounting for the fact that we actually simulate only primaries with $m_{\rm p}\ge{}5$ M$_\odot$, neglecting lower mass stars. Finally, $f_{\rm bin}$ accounts for the fact that we simulate only binary systems, whereas a fraction of stars are single. Here we assume that 50 per cent of stars are in binaries, thus $f_{\rm bin}=0.5$. We note that  $f_{\rm bin}$ is only a scale factor and our results can be rescaled to a different  $f_{\rm bin}$  a posteriori. We notice that $N_{\rm BSE}\,{}f_{\rm corr}\,{}f_{\rm bin}/M_{\rm BSE}$ is (by definition) the number of merging BHBs per unit stellar mass at a given metallicity. 

With this procedure, we associate to each Illustris' particle a number $n_{\rm BHB}$ of randomly selected merging BHBs whose progenitors have  metallicity $Z\simeq{}Z_{\rm Ill}$.

We then estimate the look-back time of the merger ($t_{\rm merg}$) of each BHB in the randomly selected sample as
\begin{equation}
t_{\rm merg}=t_{\rm form}-t_{\rm delay},
\end{equation}
where $t_{\rm delay}$ is the time between the formation of the progenitor stellar binary and the merger of the BHB, and $t_{\rm form}$ is the look back time at which the Illustris' particle has formed, calculated as
\begin{equation}
t_{\rm form}=\frac{1}{H_0}\int_0^{z_{\rm Ill}}\frac{1}{(1+z)\,{}\left[\Omega{}_M\,{}(1+z)^3+\Omega{}_\Lambda{}\right]^{1/2}}{\rm d}z,
\end{equation}
where the cosmological parameters are set to WMAP-9 values (for consistency with the Illustris) and $z_{\rm Ill}$ is the formation redshift of the Illustris' particle.

According to this definition, $t_{\rm merg}$ is also a look back time: it tells us how far away from us the BHB merged. For our analysis, we consider only BHBs with $t_{\rm merg}\ge{}0$, i.e. we do not consider BHBs that will merge in the future.

We repeat the same procedure for each of the  six simulation sets in Table~\ref{tab:table1} and we obtain six different models of the cosmic BHB merger evolution.

\section{Results}\label{sec:results} 

\begin{figure}
\center{{
\epsfig{figure=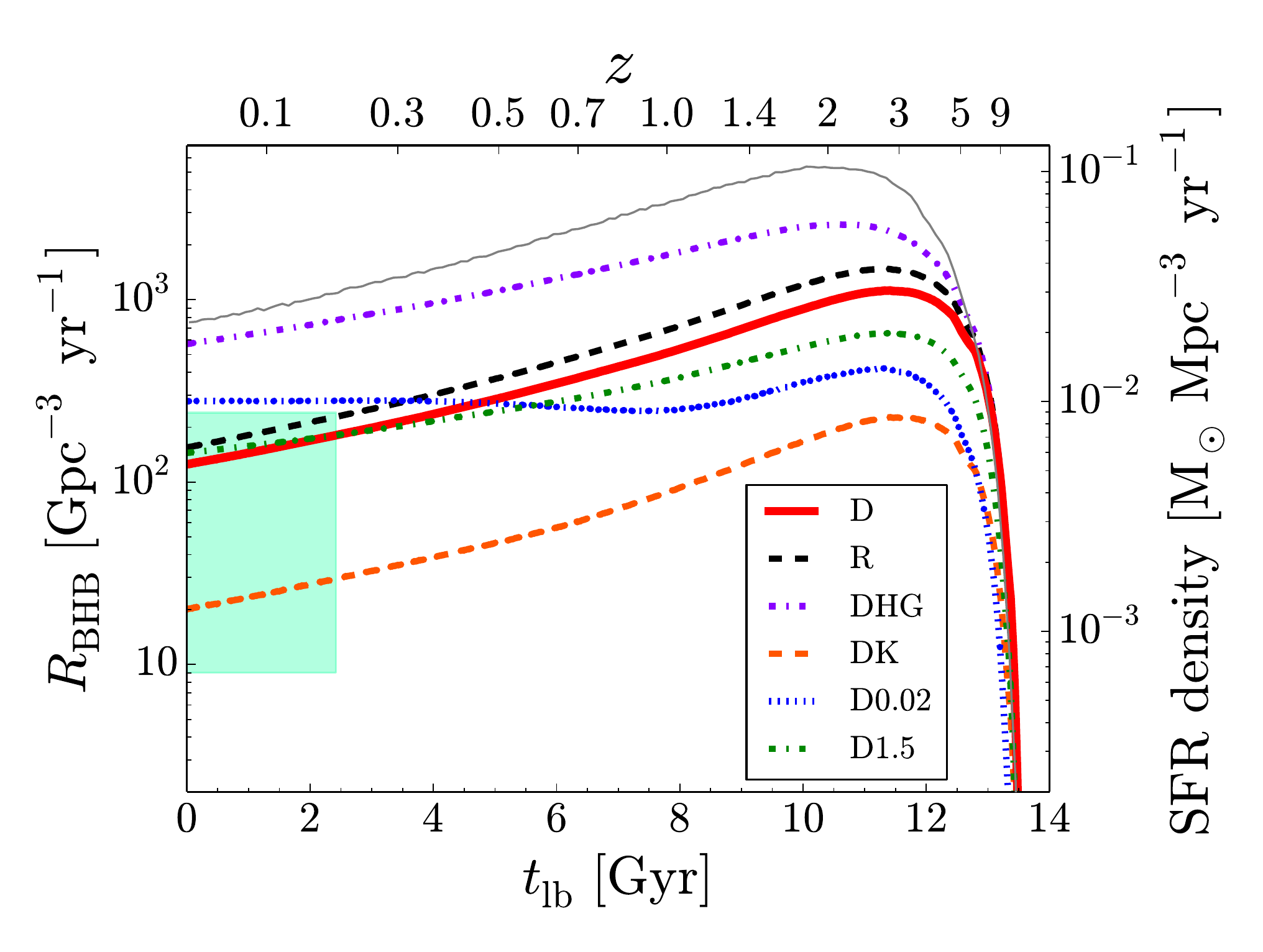,width=9cm} 
}}
\caption{\label{fig:fig1}
Left $y-$axis: cosmic merger rate density of BHBs ($R_{\rm BHB}$) in the comoving frame, as a function of the look-back time $t_{\rm lb}$ (bottom $x$ axis) and of the redshift $z$ (top $x$ axis) in our models. Red solid line: D (fiducial model);  black dashed line: R; violet dash-dot line: DHG; orange dashed line: DK; blue dotted line: D0.02;  green dash-dot line: D1.5. Green shaded area: BHB merger rate inferred from LIGO detections \citep{abbott2016cO1}. Right $y-$axis: cosmic SFR density from the Illustris (grey thin solid line), as a function of the look-back time $t_{\rm lb}$ (bottom $x$ axis) and of the redshift $z$ (top $x$ axis). 
}
\end{figure}

\subsection{Merger rate}\label{sec:mergrate}
Figure~\ref{fig:fig1} shows the cosmic BHB merger rate density ($R_{\rm BHB}$) in the comoving frame, derived from our simulations. To obtain the BHB merger rate shown in this Figure, we extracted the number of BHB mergers ($N_{\rm BHB}$) per time bin (each bin spanned over $\Delta{}t=10$ Myr from $z\sim{}16$ to $z=0$) and then we did the following simple conversion:
\begin{equation}
R_{\rm BHB}=N_{\rm BHB}\,{}\left(\frac{l_{\rm box}}{{\rm Gpc}}\right)^{-3}\,{}\,{}\left(\frac{\Delta{}t}{\rm yr}\right)^{-1},
\end{equation}
where $l_{\rm box}=106.5$ Mpc is the size of the Illustris box (in the comoving frame) and $\Delta{}t$ is the size of the time bin ($\Delta{}t=$10 Myr).

\begin{table}
\begin{center}
\caption{\label{tab:table2}
Comoving BHB merger-rate density at redshift $z=0$ and $z=0.2$ (corresponding to $t_{\rm lb}=0$ and 2.43 Gyr, respectively).} 
 \leavevmode
\begin{tabular}[!h]{ccc}
\hline
Name &  $R_{\rm BHB} (z=0)$ & $R_{\rm BHB}(z=0.2)$\\ 
     & [Gpc$^{-3}$ yr$^{-1}$] & [Gpc$^{-3}$ yr$^{-1}$]\\
\hline
D  & 125 & 181\\
R  & 155 & 228 \\
DHG & 572 & 772 \\
DK  & 20 & 29 \\
D1.5  & 145 & 181 \\
D0.02  & 278 & 279\\
\noalign{\vspace{0.1cm}}
\hline
\end{tabular}
\begin{flushleft}
\footnotesize{Column 1: model name; column 2: present-time BHB merger rate density; column 3: BHB merger rate density at $z=0.2$.}
\end{flushleft}
\end{center}
\end{table}


From Fig.~\ref{fig:fig1} it is apparent that the overall behaviour of the merger rate is the same for all considered BHB models, with the partial exception of D0.02 (see Table~\ref{tab:table1} for details about the models). The behaviour of the merger rate density as a function of time depends only on the SFR (given by the Illustris and thus common to all BHB models) and on the delay between the formation time of a stellar binary and the merger time of the BHB born from the stellar binary (which depends on the BHB model). 

The shape of the merger rate density in Fig.~\ref{fig:fig1} resembles the one of the cosmic SFR density (e.g. \citealt{madaudickinson2014}) with a peak at $t_{\rm lb}=11.29$ Gyr  (i.e. $z\sim{}2.7$). The decrease of the merger rate approaching $z=0$ is more gentle than the decrease of the SFR density, because of BHBs that formed at high redshift but merge with a delay of several Gyr (see the next section).

The main difference between the considered BHB models is the normalization of the merger rate density, which depends on the BHB merger efficiency  (see Table~\ref{tab:table2} for details). In particular, the present-time merger rate density ranges from $R_{\rm BHB}\sim{}125$ Gpc$^{-3}$ yr$^{-1}$ to $R_{\rm BHB}\sim{}155$ Gpc$^{-3}$ yr$^{-1}$ in models D, D1.5 and R. Model D0.02 results in a factor of two larger rate ($R_{\rm BHB}\sim{}280$  Gpc$^{-3}$ yr$^{-1}$). Finally, the rate is much higher  ($R_{\rm BHB}\sim{}570$  Gpc$^{-3}$ yr$^{-1}$) in model DHG and much lower ($R_{\rm BHB}\sim{}20$  Gpc$^{-3}$ yr$^{-1}$) in model DK (see Table~\ref{tab:table2}).

The BHB merger rate density inferred from the first LIGO observations (O1 run) is $R_{\rm BHB}=9-240$ Gpc$^{-3}$ yr$^{-1}$ 
\citep{abbott2016cO1}.  While this paper was being reviewed, the inferred rate was updated to  $R_{\rm BHB}=12-213$ Gpc$^{-3}$ yr$^{-1}$, based on the first results of the O2 run \citep{abbott2017}.  Thus, the present-day merger rate density of models D, R, DK, and D1.5 are consistent with observations, while D0.02 is slightly above  the observed range  and DHG gives a much higher rate. Thus, models in which HG stars can survive a CE phase (DHG) are not consistent with the observed merger rate, unless natal kicks are much higher than assumed.

\begin{figure}
\center{{
\epsfig{figure=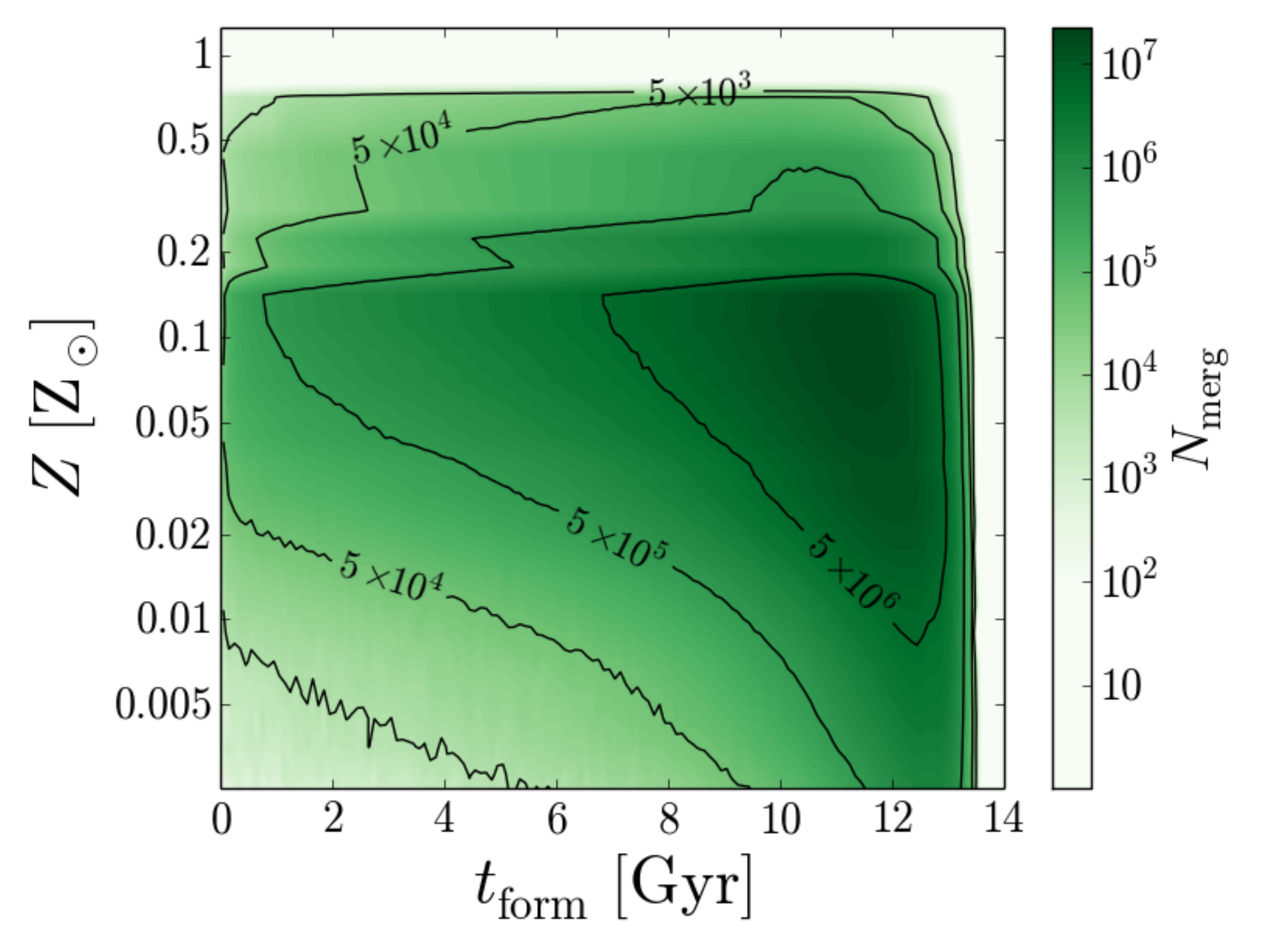,width=8.5cm} 
\epsfig{figure=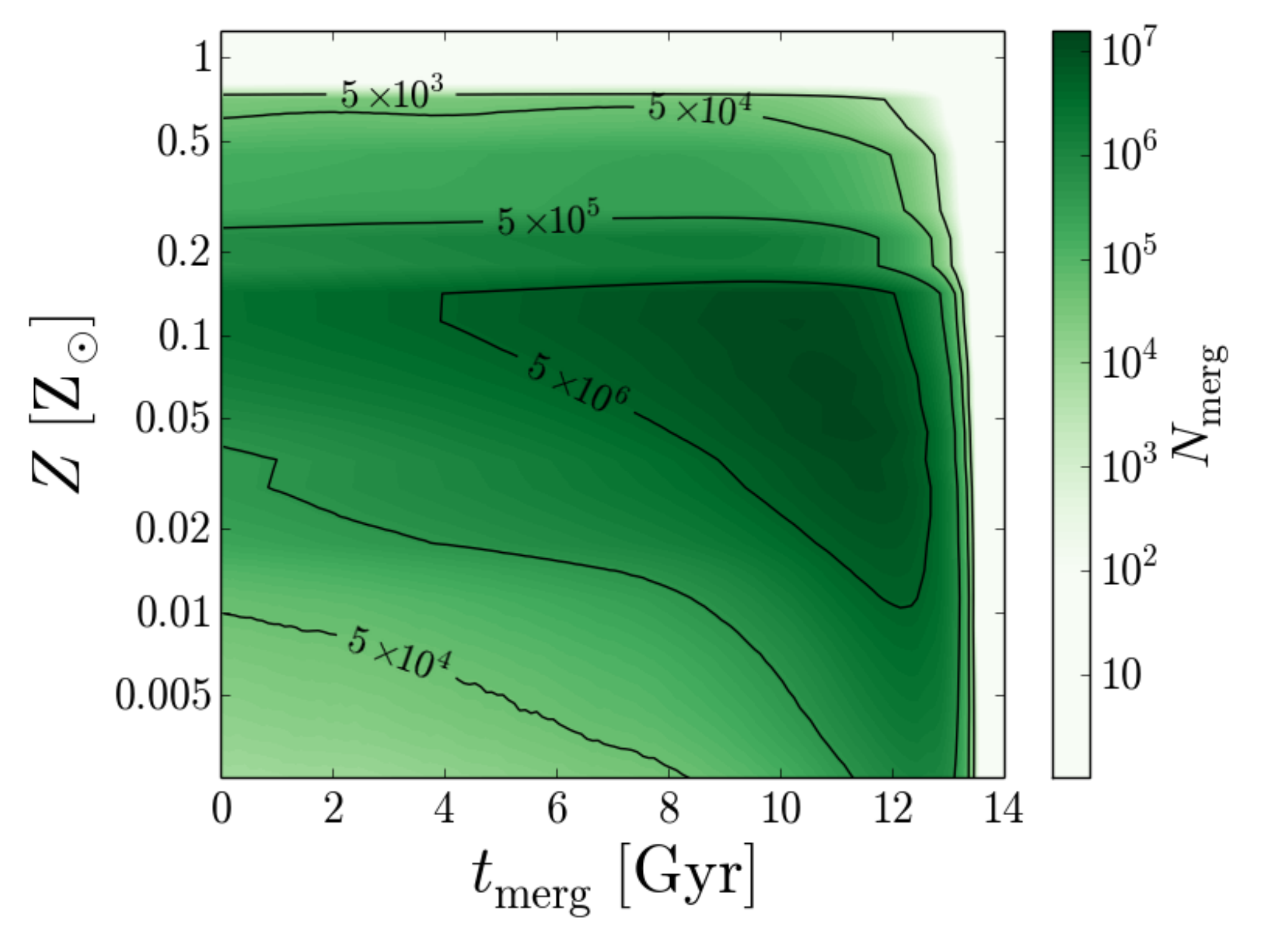,width=8.5cm} 
\caption{\label{fig:fig2}
 Metallicity of progenitors of merging BHBs in the fiducial model (D). Upper panel: metallicity versus formation time of the stellar progenitors ($t_{\rm form}$). Lower panel: metallicity of the stellar progenitors versus merger time of the BHBs ($t_{\rm merg}$).  Both $t_{\rm form}$ and $t_{\rm merg}$ are expressed as look-back time. The colour-coded map (in logarithmic scale) indicates the number of merging BHBs per cell. The black lines are isocontours enclosing a number of merging BHBs ranging from $5\times{}10^3$  to $5\times{}10^6$ (as indicated by the black labels).
}}}
\end{figure}

\begin{figure}
\center{{
\epsfig{figure=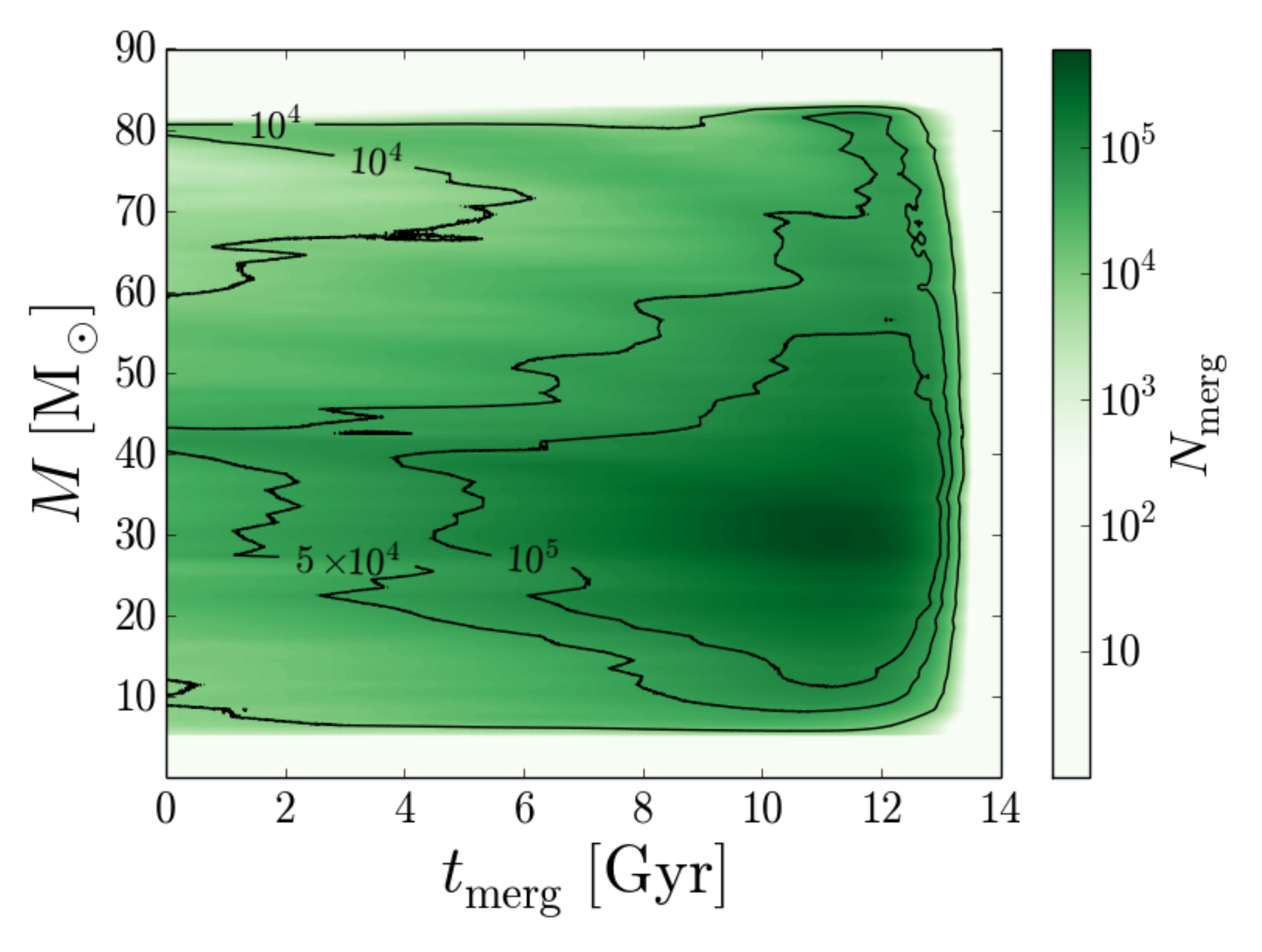,width=8.5cm} 
\epsfig{figure=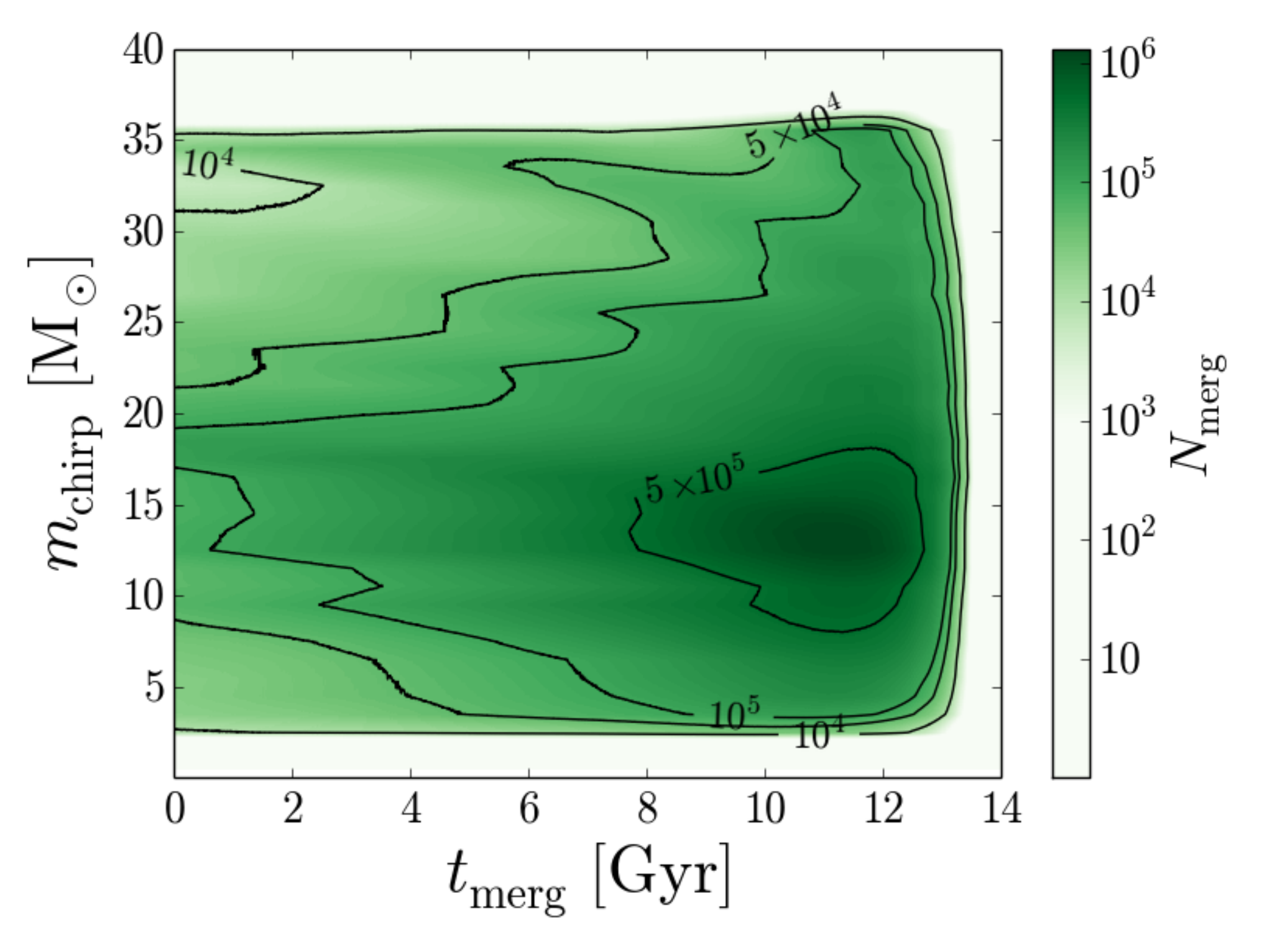,width=8.5cm} 
\caption{\label{fig:fig3}
 Upper (lower) panel: total mass (chirp mass) of merging BHBs as a function of $t_{\rm merg}$ in the fiducial model. $t_{\rm merg}$ is expressed as look-back time. The colour-coded map (in logarithmic scale) indicates the number of merging BHBs per cell. The black lines are isocontours enclosing a number of merging BHBs ranging from $5\times{}10^3$  to $5\times{}10^5$ (as indicated by the black labels).
}}}
\end{figure}

Natal kicks have a strong impact on the BHB merger rate: $R_{\rm BHB}$ is a factor of $\sim{}6$ lower in run DK than in run D, which differ only by the kick prescription. In run D the magnitude of the kick depends on the amount of fallback. In run DK all BHs receive a natal kick, drawn from the same distribution as Galactic single pulsars \citep{hobbs2005}.  
The merger rate density of both run D and DK are consistent with current observations, but future detections might be able to discriminate between such models.

Runs~D, D0.02 and D1.5 differ by the choice of the $\alpha{}$ and $\lambda{}$ CE parameters. Unlike the other considered effects (SN model, SN kicks and HG treatment), the choice of CE parameters affects not only the normalization but also the shape of the BHB merger rate density as a function of time. Since the SFR history is the same for all models, this difference indicates that models with different CE parameters have also different distributions for the delay time.

The effect of the choice of $\alpha{}\,{}\lambda{}$ is much more important at high redshift than at low redshift. At $z<0.3$, the difference between runs D and D1.5 is negligible, while the difference between run D and D0.02 is about a factor of two.

Finally, runs R and D have  similar merger rates (within a factor of 1.3).  This indicates that the choice of the core-collapse SN model (rapid or delayed) does not affect the BHB merger rate significantly. In the following, we will consider run D as our fiducial model.

\subsection{Formation time, progenitor's metallicity and BHB masses}\label{sec:sec3.2}
In this section, we discuss the main properties of merging BHBs in the Illustris simulation. We consider only our fiducial run~D. 

 Figure~\ref{fig:fig2} maps the metallicity of the stellar progenitors of merging BHBs. In the upper panel the metallicity is plotted against the formation time of the stellar progenitors, while the lower panel shows the metallicity versus the merger time of BHBs. From the comparison between the two panels, it is apparent that a large fraction of metal-poor systems which formed at high redshift merge at relatively low redshift with a long delay time. For example, $\sim{}2\times{}10^6$ BHBs with progenitor metallicity $Z\sim{}0.1$ Z$_\odot$ merge at redshift $z\sim{}0$ in the simulation, but only $\sim{}5\times{}10^4$ of them form at redshift $z\sim{}0$. This implies that a significant number of merging BHBs visible in the LIGO instrumental horizon  were born in the high-redshift Universe and possibly in a metal-poor environment.

Figure~\ref{fig:fig2} also shows that the stellar progenitors of BHBs  have all possible metallicities ranging from $Z\sim{}0$ up to $\sim{}0.7\,{}Z_{\odot}$. 
The most common metallicity of BHBs merging at low redshift is $0.05\lesssim{}Z/{\rm Z}_{\odot}\lesssim{}0.2$, i.e. significantly sub-solar. This result comes from a combination of two factors. Firstly, relatively metal-poor stars form efficiently even at $z=0$, as expected from the mass-metallicity relation. Secondly, many BHBs merging at $z=0$ formed at high-redshift, where low metallicity was more common. Mergers associated with solar or super-solar metallicity are strongly suppressed in our models, because stellar radii are larger at higher metallicity, causing early mergers of massive stars before they become BHBs.

At higher redshift, the percentage of merging BHBs born in metal-poor environments increases, and the contribution of lower metallicities becomes more important. However, we stress that even at $t_{\rm merg}\gtrsim{}11$ Gyr there is a non-negligible fraction of systems with metallicity $>0.1$ Z$_\odot$.

Figure~\ref{fig:fig3} shows the behaviour of the total mass ($M=m_{\rm p}+m_{\rm s}$, where $m_{\rm p}$ and $m_{\rm s}$ are the mass of the primary and secondary BH, respectively) and of the chirp mass ($m_{\rm chirp}=m_{\rm p}^{3/5}\,{}m_{\rm s}^{3/5}\,{}M^{-1/5}$) as a function of the merger time.  The distribution of total masses (chirp masses) peaks at   $20\le{}M/{\rm M}_\odot\le{}45$ ($8\le{}m_{\rm chirp}/{\rm M}_\odot\le{}20$), but we find merging systems with total masses (chirp masses) ranging from $\sim{}6$ to $\sim{}82$ M$_\odot$ ($\sim{}3$ to $\sim{}35$ M$_\odot$). From this Figure it is apparent that there is nearly no dependence of the BHB mass on the merging time. This is primarily a consequence of the broad distribution of delay times  (see next Section).


\begin{figure}
\center{{
\epsfig{figure=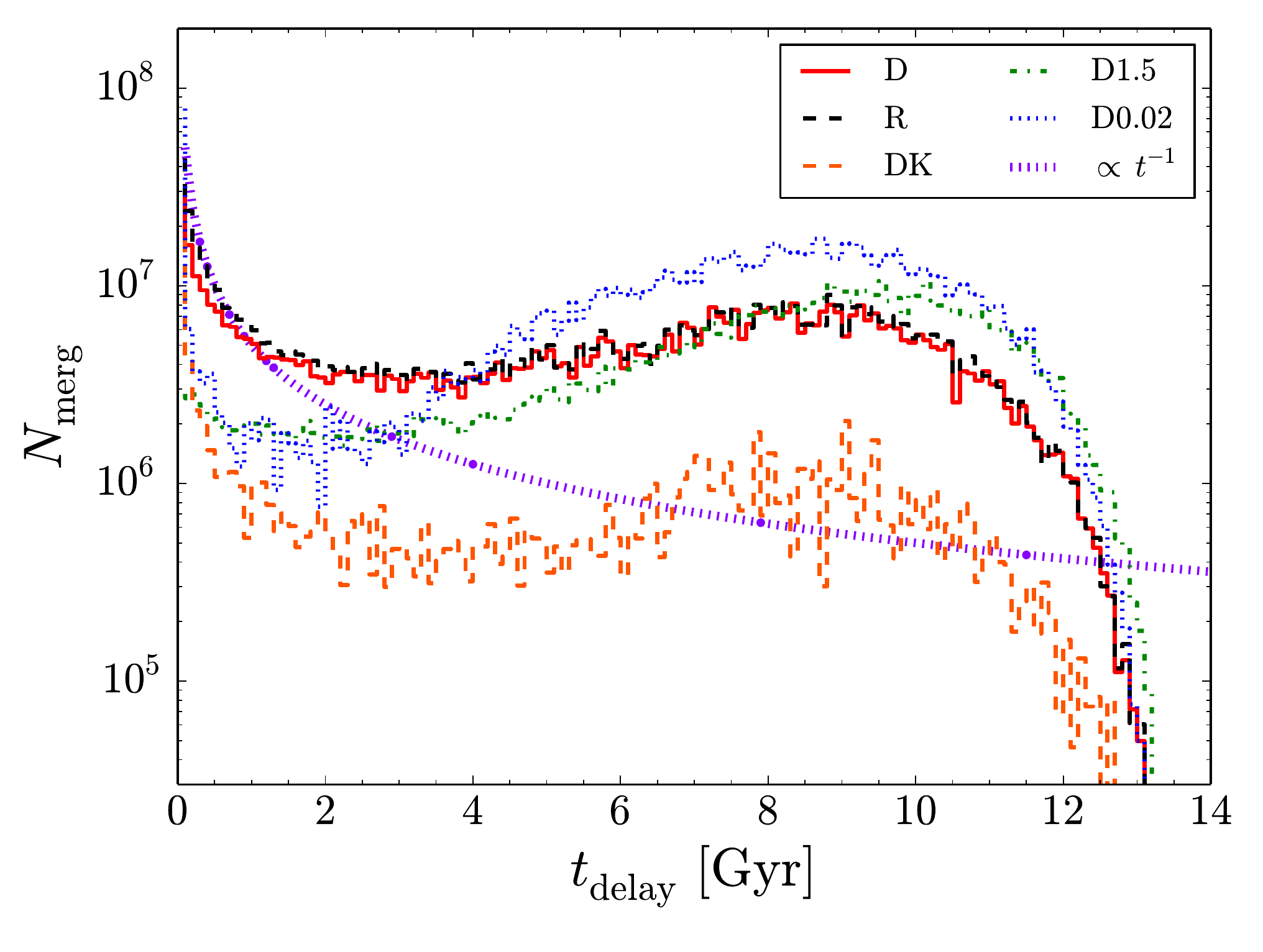,width=8cm} 
}}
\caption{\label{fig:fig4}
Distribution of the delay time ($t_{\rm delay}$, estimated as the time elapsed between the formation of a stellar binary and the merger of the BHB formed from this stellar binary) for the simulated BHBs in runs~D (red solid line), R (black dashed line), DK (orange dashed line), D1.5 (green dash-dot line) and D0.02 (blue dotted line). Only BHBs merging within the mass dependent instrumental horizon of LIGO are shown.  Purple dotted line: ${\rm d}N/{\rm d}t\propto{}t^{-1}$. Note that (unlike $t_{\rm merg}$ and $t_{\rm form}$) $t_{\rm delay}$ is not a look-back time. The value $N_{\rm merg}$ on the $y$ axis is the number of simulated BHBs per time bin ($\Delta{}t=100$ Myr). 
}
\end{figure}

\subsection{BHBs merging within the LIGO horizon}
In this section, we focus only on simulated BHBs that merge within the LIGO instrumental horizon, defined as the luminosity distance at which GWs from a face-on, equal-mass, overhead binary would be detected with signal-to-noise ratio of 8 \citep{abbott2016bASTRO}. 
To account for the dependence of the instrumental horizon on the BHB mass, we use the curve reported in Fig. 4 of \cite{abbott2016bASTRO} for the 2015-2016 LIGO sensitivity (left-hand panel). 

In order to extract from our simulations all BHBs merging inside the LIGO instrumental horizon, we check whether the luminosity distance at the time of merger is smaller than the instrumental horizon for a BHB with the same total mass, as given in  Fig. 4 of \cite{abbott2016bASTRO}. 



\begin{figure}
\center{{
\epsfig{figure=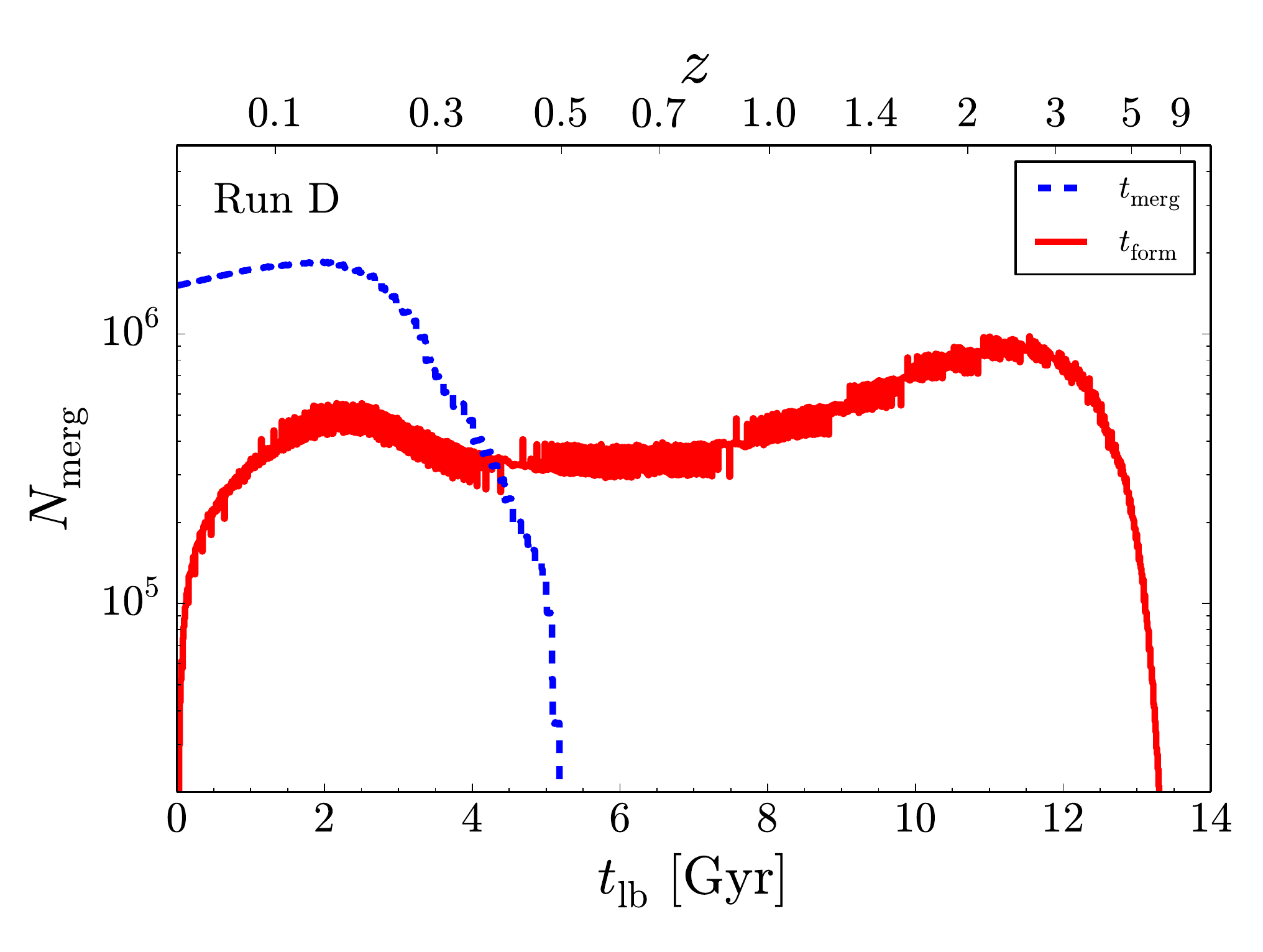,width=8cm} 
}}
\caption{\label{fig:fig5}
Red solid line: distribution of formation time $t_{\rm form}$ for the simulated BHBs in the fiducial model (run D). Blue  dashed line: distribution of merger time $t_{\rm merg}$ for the simulated BHBs in the fiducial model. Only BHBs merging within the LIGO instrumental horizon are shown. Bottom $x$ axis: $t_{\rm form}$ and $t_{\rm merg}$ expressed as look-back time. Top $x$ axis: $t_{\rm form}$and $t_{\rm merg}$ expressed as redshift. The value $N_{\rm merg}$ on the $y$ axis is the number of simulated BHBs per time bin ($\Delta{}t=10$ Myr).
}
\end{figure}

In this section we also compare the properties of our fiducial model with the other runs. Figure~\ref{fig:fig4} shows the delay time distribution for the BHB merging within the LIGO horizon. This distribution depends only on the \textsc{BSE} models and is not affected by the cosmological simulation. In run~D, the delay distribution matches the behaviour ${\rm d}N/{\rm d}t\propto{}t^{-1}$ found in previous studies \citep{belczynski2016,lamberts2016}, but only for $t_{\rm delay}\le{}2$ Gyr. For longer delay times, the distribution flattens considerably: it becomes nearly independent of time.  This explains why a large fraction of metal-poor systems formed at high redshift merge within the LIGO horizon (see Fig.~\ref{fig:fig2} and Section~\ref{sec:sec3.2}).  Runs~R  (adopting the rapid SN model) and DK  (assuming large BH kicks) behave exactly the same as run D. 

In contrast, the delay time distributions of runs~D1.5 and especially D0.02 (which differ from run~D only for the CE parameters) are significantly different. They decrease more steeply than $\propto{}t^{-1}$ at short delay time. This produces a much lower number of BHB mergers with delay time $0.01\le{}t_{\rm delay}/{\rm Gyr}\le{}4$. For $t_{\rm delay}>4$ Gyr, the number of BHB mergers in runs D1.5 and D0.02 becomes significantly higher than that of run~D. This difference in the distribution of delay times explains why the BHB merger rate density in run~D is higher (lower) at high (low) redshift than that of runs~D1.5 and D0.02.

Figure~\ref{fig:fig5} shows the distribution of formation times $t_{\rm form}$ and the merger times $t_{\rm merg}$ of BHBs that merge within the LIGO horizon in our fiducial model. In Fig.~\ref{fig:fig5}, the merger time peaks at $z\sim{}0.1-0.2$. This results from the convolution between the LIGO horizon (which depends on the BHB mass) and the increase of the merger rate as the redshift increases (see Fig.~\ref{fig:fig1}). Interestingly,   the estimated redshift of GW150914 and GW151226 is $z\sim{}0.1$, while $z\sim{}0.2$ is the redshift of the candidate event (LVT151012) and of the most recent detection (GW170104). 


The distribution of  $t_{\rm form}$ and $t_{\rm merg}$  in the other models is similar to the one shown in Fig.~\ref{fig:fig5}. The main differences arise from the distribution of delay times. For this reason, the distribution of formation times in run D0.02 has a much higher peak at high redshift than that of run~D.

\begin{figure}
\center{{
\epsfig{figure=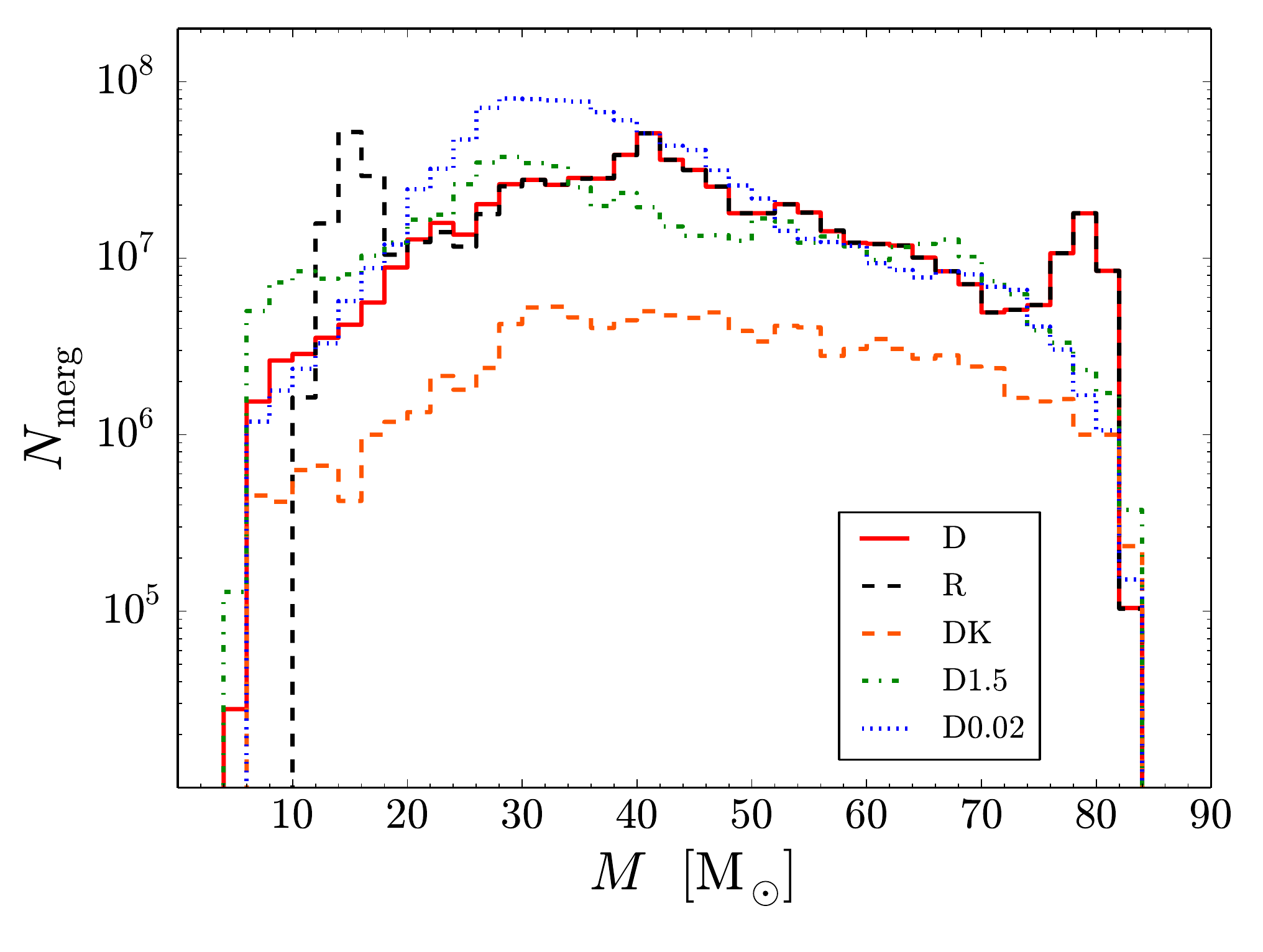,width=8cm} 
\epsfig{figure=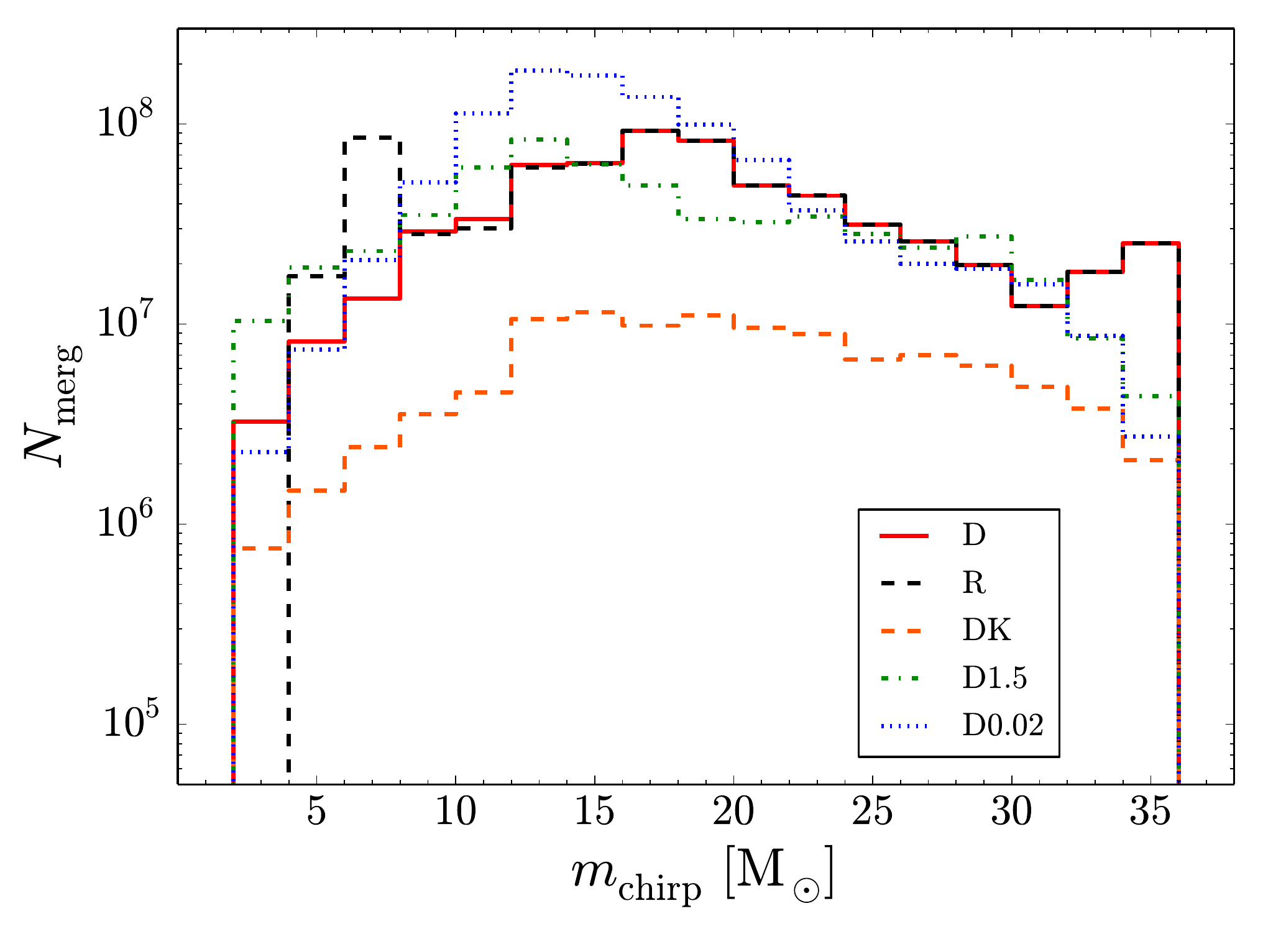,width=8cm} 
}}
\caption{\label{fig:fig6}
Distribution of total masses (upper panel) and chirp masses (lower panel) of the simulated BHBs merging within the LIGO horizon. Red solid line: run~D; black dashed line: R; orange dashed line: DK; green dash-dot line: D1.5; blue dotted line: D0.02. The value $N_{\rm merg}$ on the $y$ axis is the number of simulated BHBs per mass bin ($\Delta{}m=2$ M$_\odot$ for both the chirp and the total mass). }
\end{figure}

Figure~\ref{fig:fig6} shows the distribution of total masses and chirp masses of the simulated BHBs that merge within the LIGO horizon. The chirp masses (total masses) range between 3 and 35 M$_\odot$ (6 and 82 M$_\odot$). 

Runs~D1.5 and D0.02 are significantly different from the others, with a peak at lower masses ($M\sim{}20-40$ M$_\odot$, $m_{\rm chirp}\sim{}8-20$ M$_\odot$) and a dearth of massive systems. Runs with $\alpha{}\,{}\lambda{}=0.1$ are more similar between each other. For high total masses ($M>20$ M$_\odot$) there is no appreciable difference between runs D and R (which differ for the SN model). At lower masses, run~R (assuming a rapid SN model) has a peak at $11\le{}M/{\rm M}_\odot{}\le{}17$, which is completely absent if the delayed SN model is assumed. Moreover, in run~D and in the other runs assuming a delayed SN models, BHs with mass down to 3 M$_\odot$ are allowed to form, while no BHs with mass $<5$ M$_\odot$ form in run~R.

Finally, the distribution of BHB masses in run~DK is very similar to that of run~D, except for one important detail: there is no peak for total BHB masses (chirp masses) $75<M/{\rm M}_\odot{}<82$ ($30<m_{\rm chirp}/{\rm M}_\odot{}<35$). This comes from the fact that in run~DK all BHs receive a strong natal kick (regardless of their mass), while in the other runs the kick is modulated by the fallback.

Figure~\ref{fig:fig6} is significantly affected by the fact that the LIGO horizon depends on the BHB mass. More massive binaries can be observed also if they merge at  higher redshift ($z\sim{}0.4$ if $M\sim{}50$ M$_\odot$). Thus, their contribution to Fig~\ref{fig:fig6} is enhanced with respect to that of lighter BHBs.

Future LIGO-Virgo detections will enable us to reconstruct total mass and chirp mass distributions. Hopefully, the observed distributions will be able to discriminate between different models, indicating which one captures the main physics of BHB evolution.

\begin{figure}
\center{{
\epsfig{figure=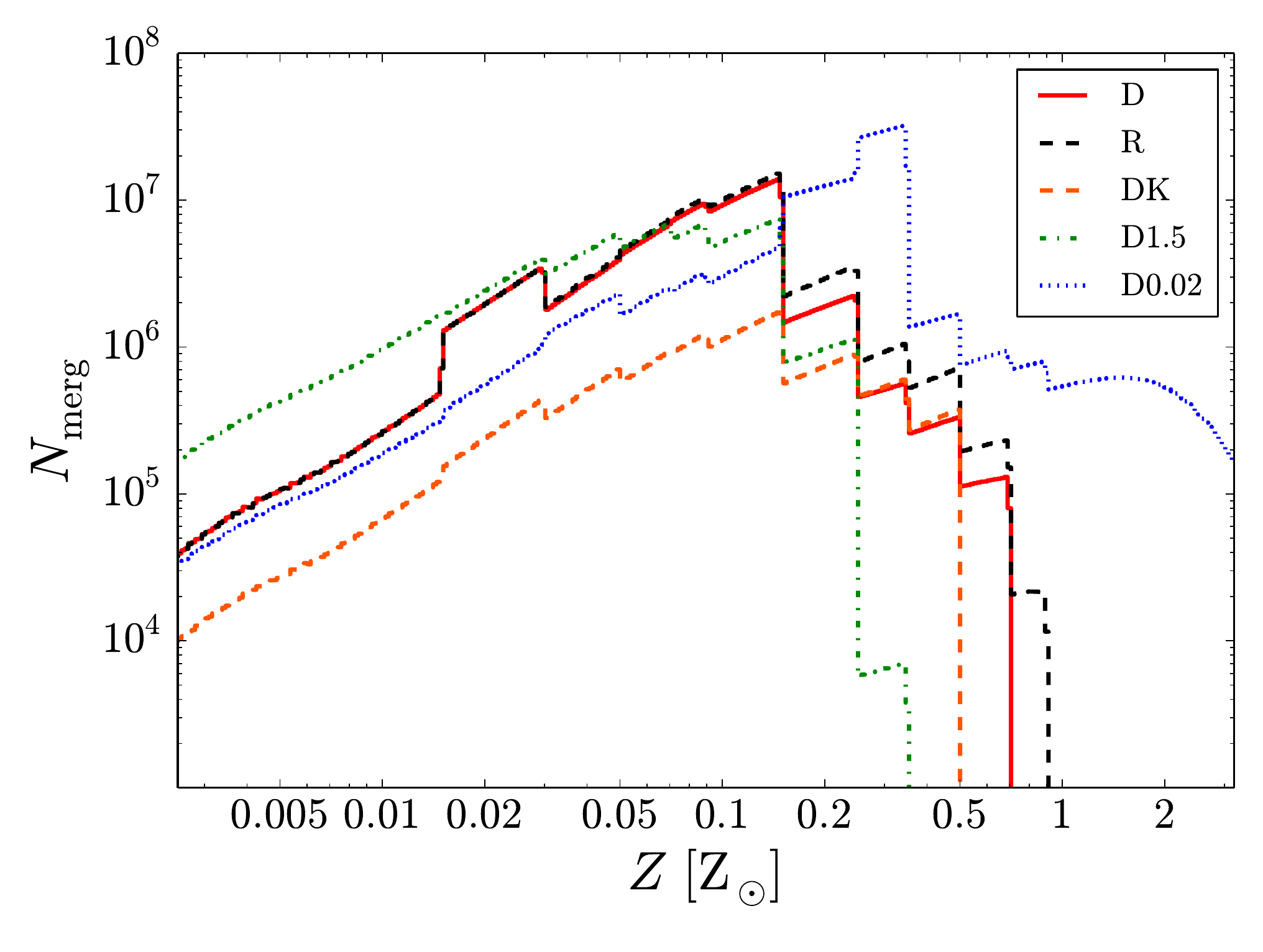,width=8cm} 
}}
\caption{\label{fig:fig7}
Distribution of metallicity ($Z$) of the stellar progenitors of the simulated BHBs that merge within the LIGO horizon.  Red solid line: run~D; black dashed line: R; orange dashed line: DK; green dash-dot line: D1.5; blue dotted line: D0.02. The value $N_{\rm merg}$ on the $y$ axis is the number of simulated BHBs per metallicity bin  ($\Delta{}\log{Z}=0.01$). The step-like features in the plot correspond to the metallicity groups in the \textsc{BSE} simulations.}
\end{figure}

Finally, Fig.~\ref{fig:fig7} shows the distribution of metallicity  of the stellar progenitors of the simulated BHBs merging within LIGO horizon. In runs D, DK and R, the metallicity of BHB progenitors peaks in the range $0.01\le{}Z/{\rm Z}_\odot\le{}0.2$. The metallicity distribution drops for $Z\gtrsim{}0.5\,{}Z_\odot{}$.  

Also in this case, runs with different CE parameters (D1.5 and D0.02) have a different trend. In run D0.02 the metallicity of BHB progenitors peaks at $0.15<Z/{\rm Z}_\odot<0.4$, significantly higher than in runs with $\alpha{}\,{}\lambda{}=0.1$. In contrast, in run D1.5 the most metal-poor systems are more efficient in producing merging BHBs than in the other runs. The step-like features clearly visible in Fig.~\ref{fig:fig7} are model artifacts, due to the fact that we simulated BHBs only in 12 metallicity bins (see Section~\ref{sec:methods}).

\begin{table}
\begin{center}
\caption{\label{tab:table3}
Percentage of simulated BHBs merging in the LIGO horizon with mass consistent with the detected events.} 
 \leavevmode
\begin{tabular}[!h]{ccccc}
\hline
Name &  GW150914 & LVT151012 & GW151226 & GW170104 \\ 
\hline
D      & 6.0\% & 9.9\%  & 3.4\% & 12.1\% \\
R      & 5.3\% & 8.9\%  & 3.3\% & 10.8\% \\
DK     & 9.8\% & 11.0\% & 2.5\% & 12.5\% \\
D1.5   & 8.4\% & 13.1\% & 5.0\% & 10.4\% \\
D0.02  & 2.8\% & 12.9\% & 4.9\% & 5.2\%\\
\noalign{\vspace{0.1cm}}
\hline
\end{tabular}
\begin{flushleft}
\footnotesize{Column 1: model name; columns 2, 3, 4, and 5: percentage of simulated BHBs with mass consistent with GW150914, LVT151012, GW151226 and GW170104, respectively, which merge within the LIGO horizon. This percentage is calculated over all simulated BHBs that merge within the LIGO horizon.}
\end{flushleft}
\end{center}
\end{table}

\begin{figure*}
\center{{
\epsfig{figure=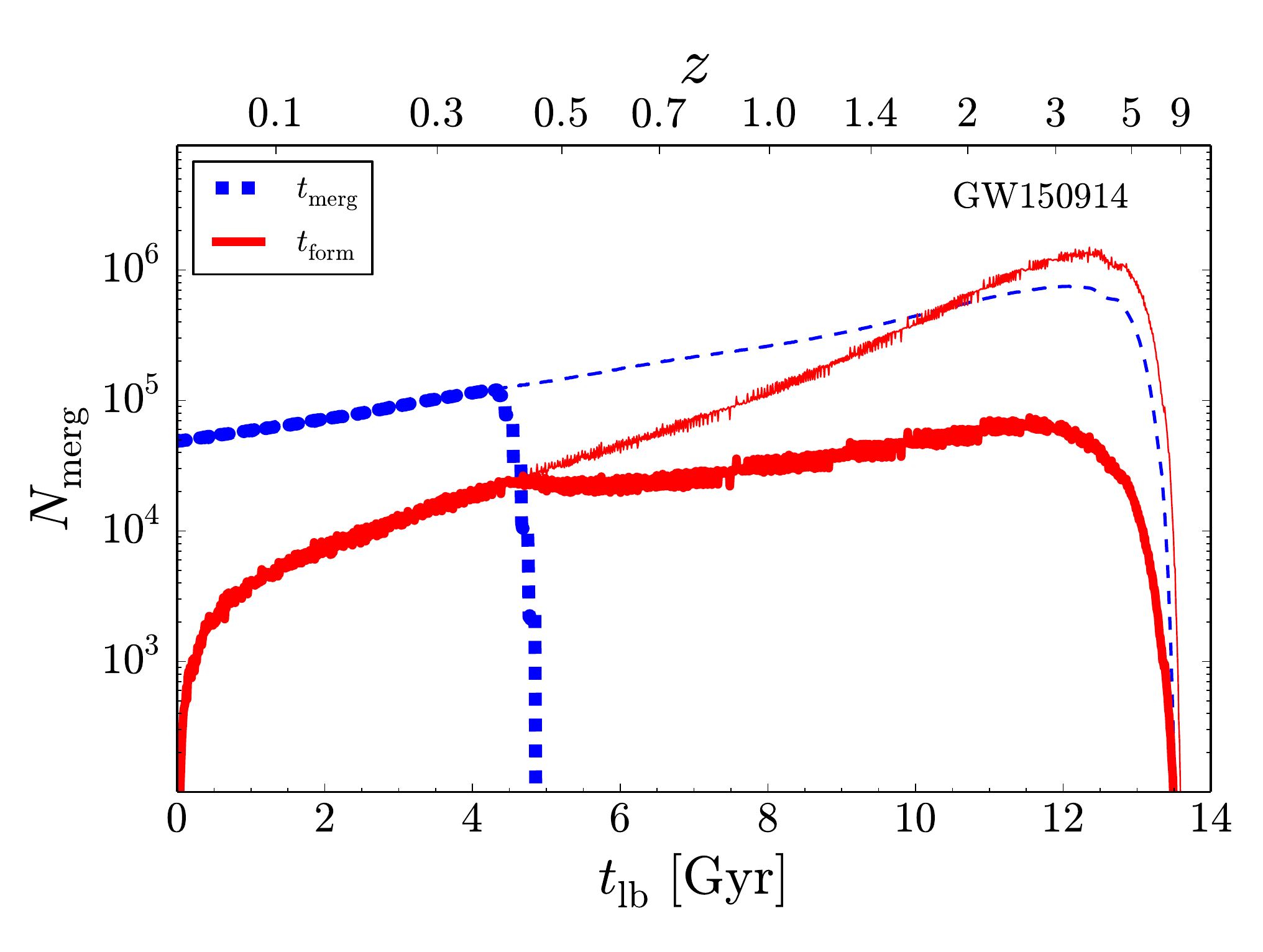,width=8cm} 
\epsfig{figure=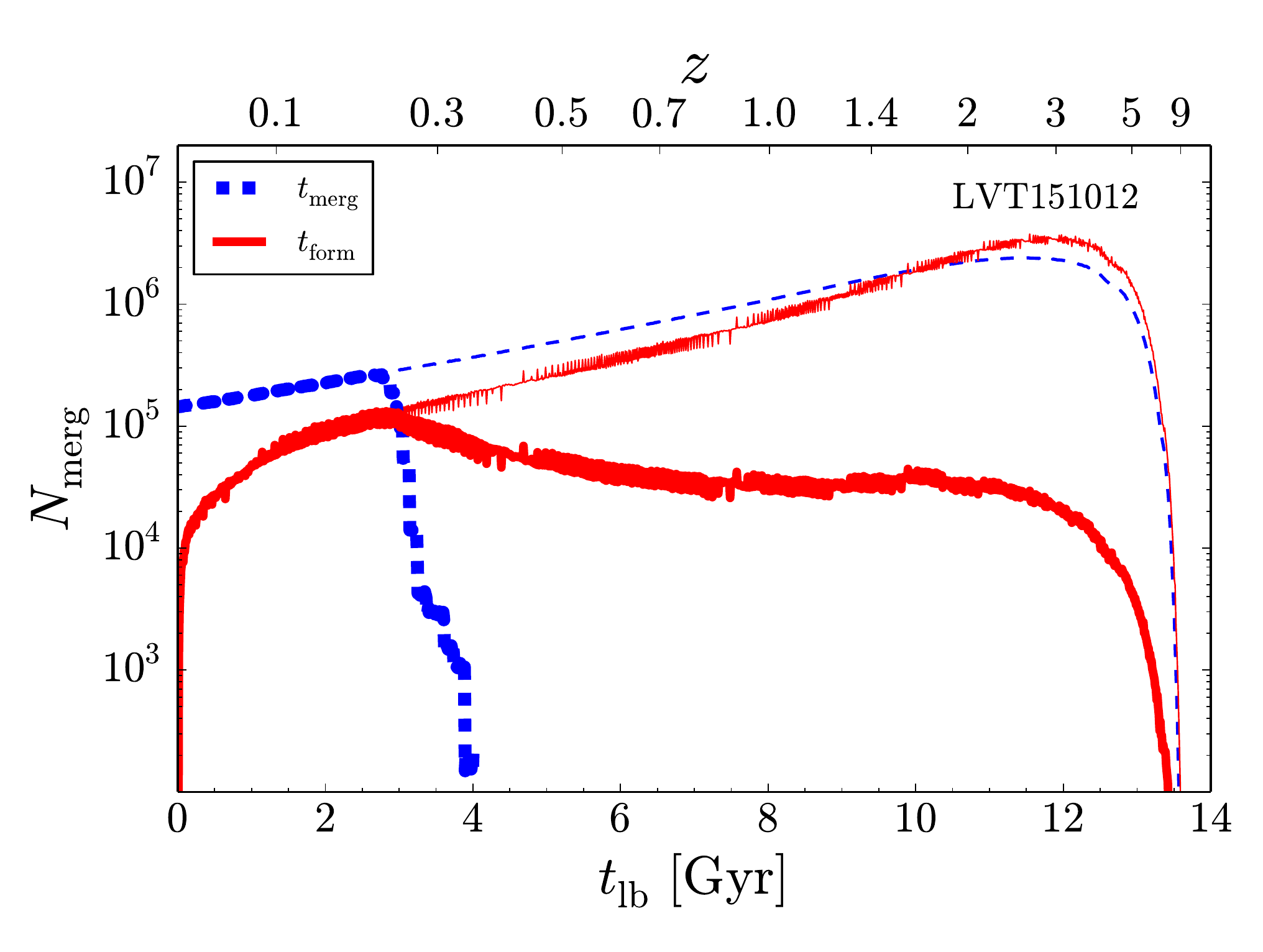,width=8cm} 
\epsfig{figure=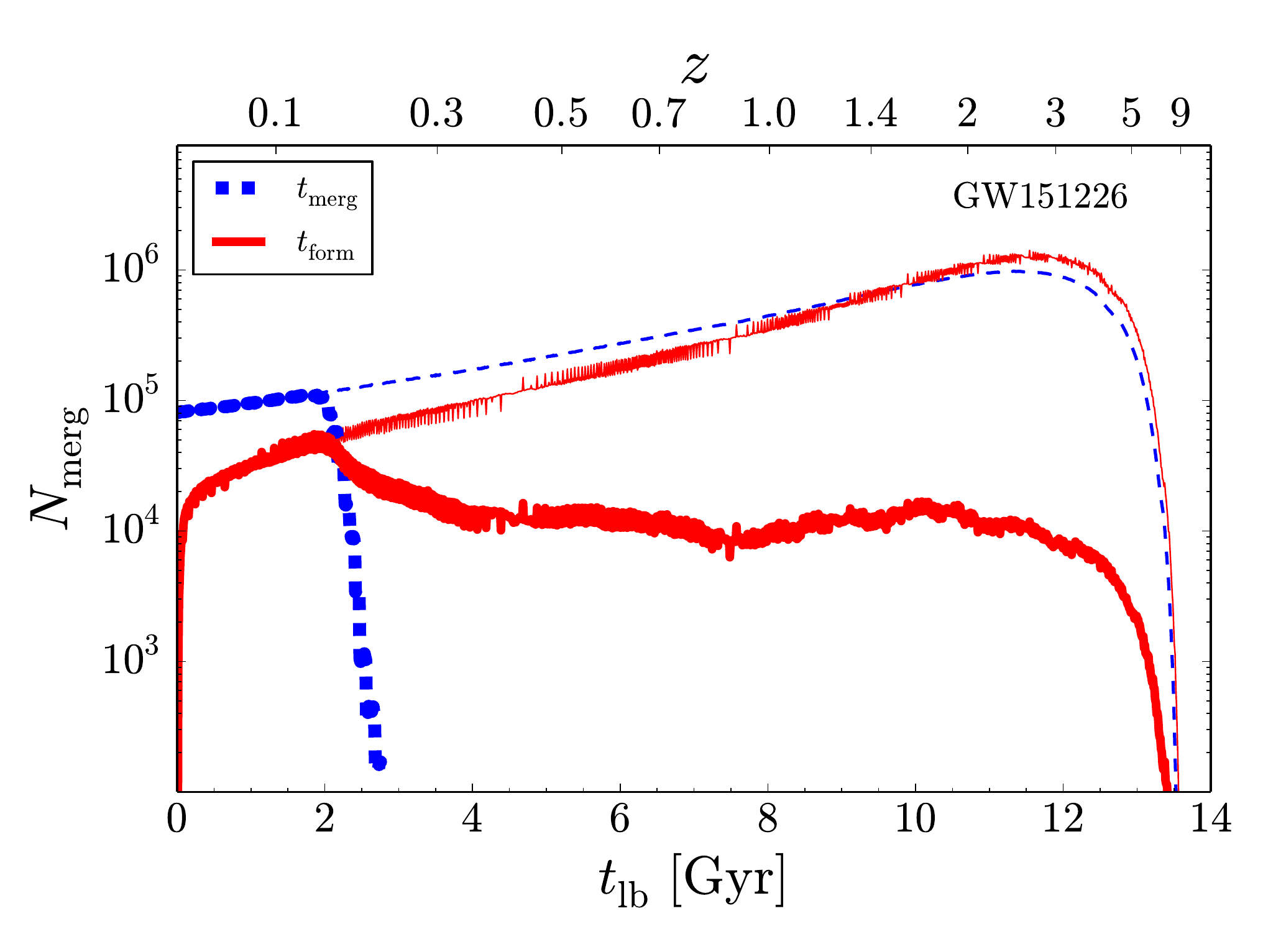,width=8cm} 
\epsfig{figure=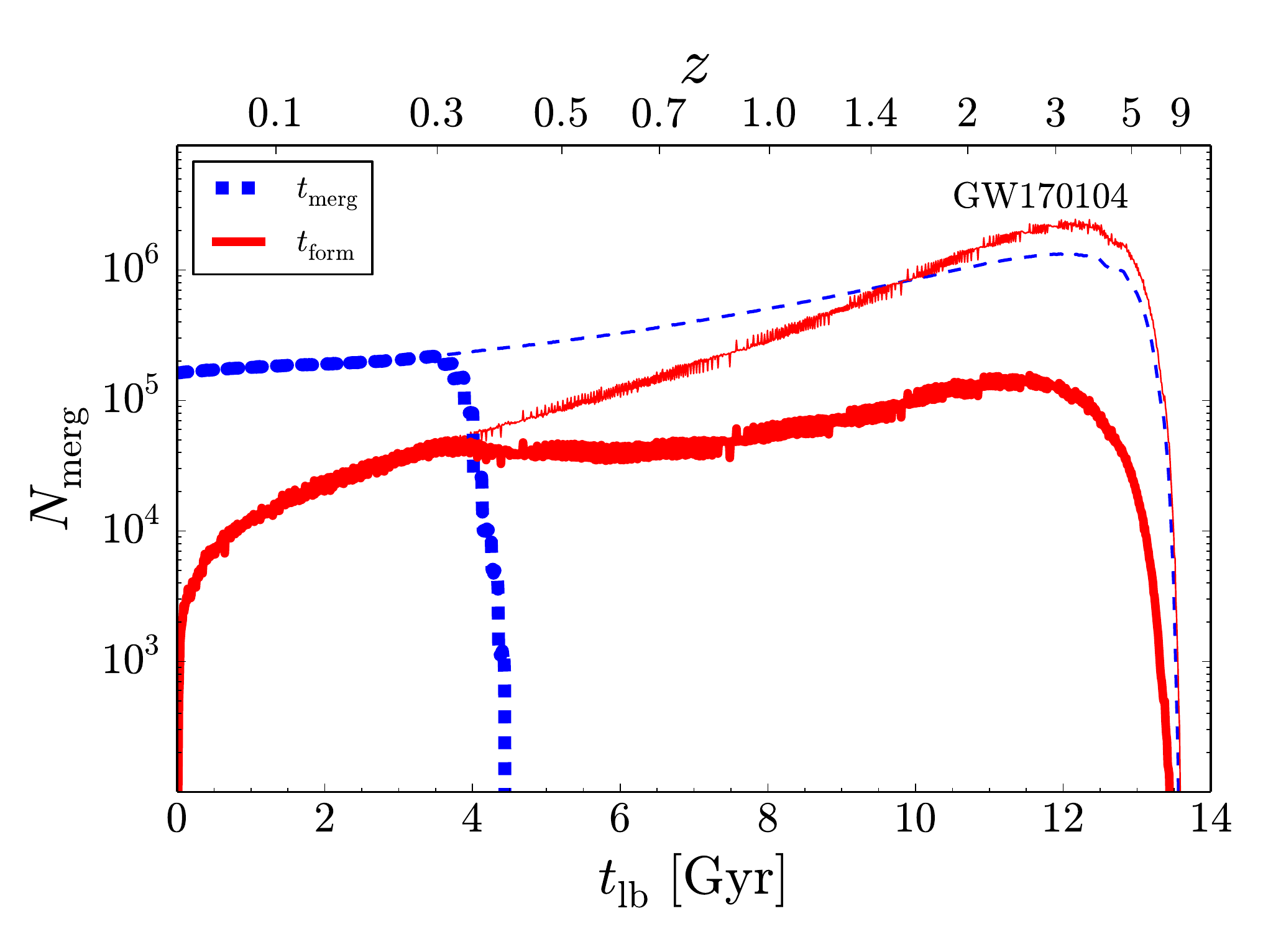,width=8cm} 
}}
\caption{\label{fig:fig8}
Distribution of the formation time ($t_{\rm form}$, red solid line) and of the merger time ($t_{\rm merg}$, blue dashed line) for the simulated BHBs matching the mass of  GW150914 (top left), LVT151012 (top right), GW151226 (bottom left) and GW170104 (bottom right) in the fiducial model (run~D). Thin lines indicate all simulated systems matching the mass of GW150914, LVT151012, GW151226, and GW170104, while thick lines indicate only systems that merge within LIGO's horizon.  Both  $t_{\rm form}$ and $t_{\rm merg}$ are expressed in look-back time (bottom $x$ axis) and redshift (top $x$ axis). The value $N_{\rm merg}$ on the $y$ axis is the number of simulated BHBs per time bin ($\Delta{}t=10$ Myr for both the merger time and the formation time). 
}
\end{figure*}

\subsection{GW150914, LVT151012, GW151226, and GW170104-like systems}
Finally, we focus on the properties of simulated systems that match the  three observed GW events, GW150914 ($m_{\rm p}=36.2^{+5.2}_{-3.8}\,{}{\rm M}_\odot$, $m_{\rm s}=29.1^{+3.7}_{-4.4}\,{}{\rm M}_\odot$, \citealt{abbott2016cO1}), GW151226 ($m_{\rm p}=14.2^{+8.3}_{-3.7}\,{}{\rm M}_\odot$, $m_{\rm s}=7.5^{+2.3}_{-2.3}\,{}{\rm M}_\odot$, \citealt{abbott2016cO1}), and GW170104 ($m_{\rm p}=31.2^{+8.4}_{-6.0}\,{}{\rm M}_\odot$, $m_{\rm s}=19.4^{+5.3}_{-5.9}\,{}{\rm M}_\odot$, \citealt{abbott2017}), and the fourth  possible signal, LVT151012 ($m_{\rm p}=23^{+18}_{-6}\,{}{\rm M}_\odot$, $m_{\rm s}=13^{+4}_{-5}\,{}{\rm M}_\odot$, \citealt{abbott2016cO1}). From our simulations, we extract all merging BHBs with masses consistent with those of the BHs associated with  GW150914, GW151226, GW170104, and possibly LVT151012 (within  90\% credible intervals, as given by \citealt{abbott2016cO1}). Table~\ref{tab:table3} shows the percentage of merging BHBs that have masses consistent with  GW150914, LVT151012, GW151226 and GW170104, if we consider only BHBs that merge within the LIGO horizon.

 BHB mergers consistent with  GW150914, LVT151012, GW151226, and GW170104  exist in all of our models. In particular,  we expect $\sim{}6$ \%, $\sim{}10$ \%, $\sim{}3$ \% and $\sim{}12$ \% of events, inside the LIGO horizon, to have mass consistent with GW150914, LVT151012, GW151226, and GW170104 respectively, in run~D (see Table~\ref{tab:table3} for the other runs). Interestingly, GW150914-like events appear to be more common than GW151226-like ones within the 2015-2016 LIGO's horizon.  Also, LVT151012-like events seem to be the most common ones within LIGO's horizon, but this is mainly due to the fact that the mass of LVT151012 is much more uncertain than that of both GW150914 and GW151226, thus more systems fall within the allowed window.



\begin{figure}
\center{{
\epsfig{figure=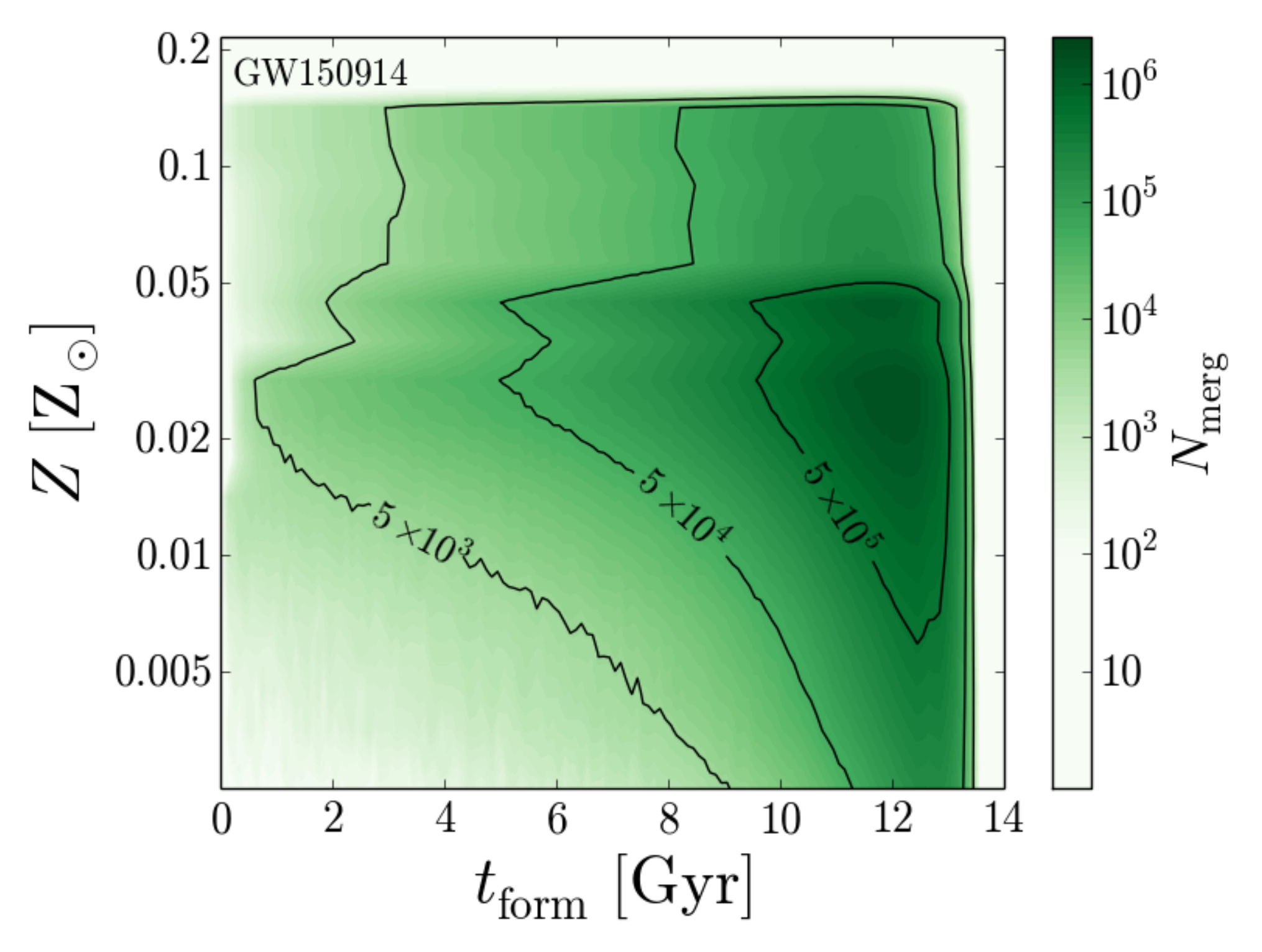,width=8cm} 
\epsfig{figure=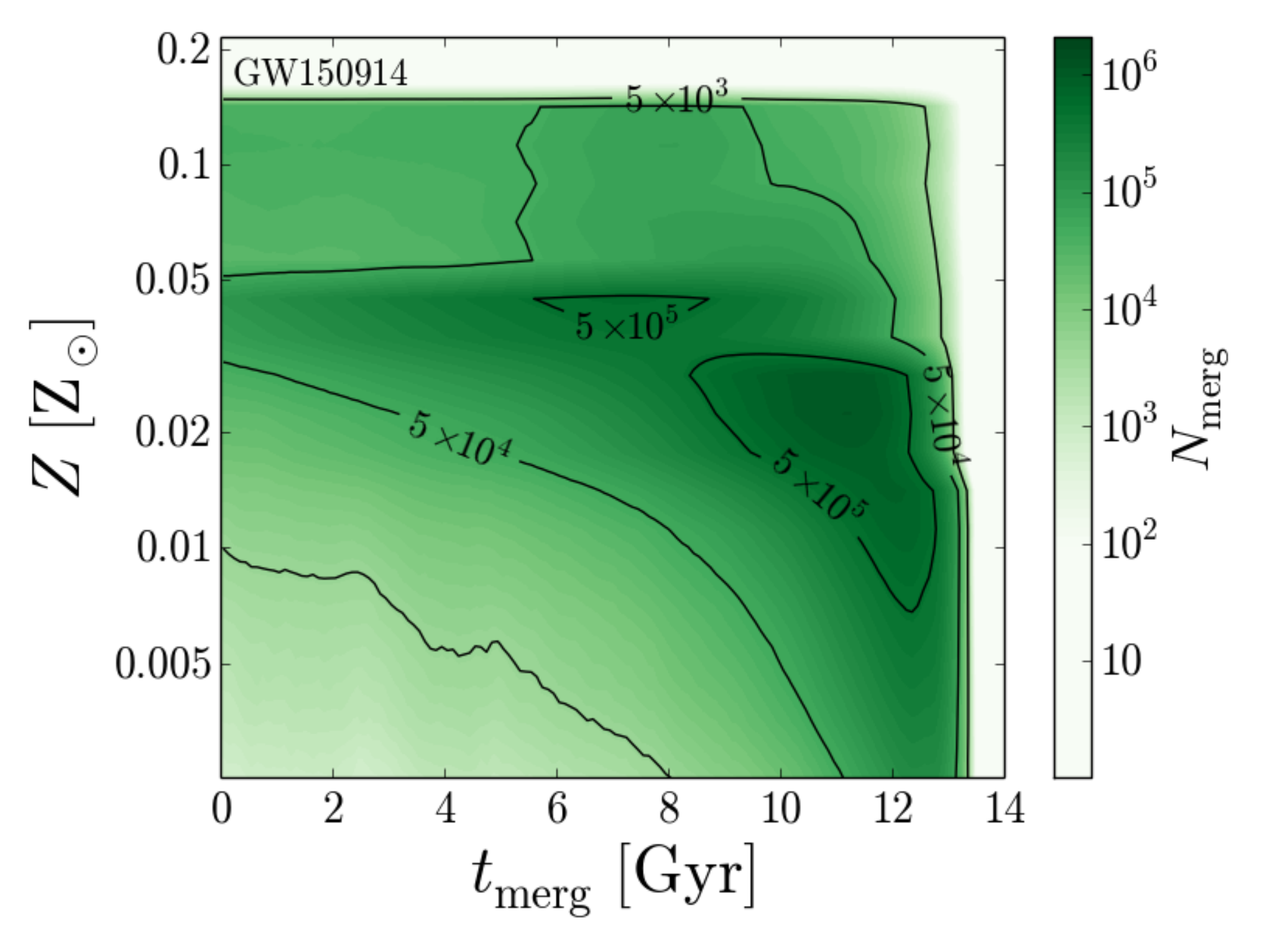,width=8cm} 
}}
\caption{\label{fig:fig9}
Metallicity of progenitors of systems matching the mass of GW150914 in the fiducial model (D). Upper panel: metallicity versus formation time of the stellar progenitors ($t_{\rm form}$). Lower panel: metallicity of the stellar progenitors versus merger time of the BHBs ($t_{\rm merg}$).  Both $t_{\rm form}$ and $t_{\rm merg}$ are expressed as look-back time. The colour-coded map (in logarithmic scale) indicates the number of merging BHBs per cell. The black lines are isocontours enclosing a number of merging BHBs ranging from $5\times{}10^3$  to $5\times{}10^5$ (as indicated by the black labels).
}
\end{figure}

Figure~\ref{fig:fig8} shows the behaviour of the merger time and the formation time for simulated systems like  GW150914, LVT151012, GW151226 and GW170104 in our fiducial model. It is apparent that the vast majority of GW150914-like systems form at high redshifts ($z>1$), with a peak at $t_{\rm form}\sim{}12-12.3$ Gyr (corresponding to $z\sim{}3.6-4.3$ in the Illustris' cosmology). Mergers also peak at high redshift ($t_{\rm merg}\sim{}12$ Gyr). However, the long delay time between formation and merger for several systems causes $t_{\rm merg}$ to decrease more gently than $t_{\rm form}$ when approaching $z=0$.

If we consider only GW150914-like systems that merge within the LIGO horizon, their formation time still peaks at high redshift ($z\sim{}2-3$), even if the slope of $t_{\rm form}$ is much less steep. This is consistent with \cite{schneider2017}, who predict that GW150914-like events form preferentially at redshift $2.4 \leq{}z\leq{}4.2$. 

In the nearby Universe, we expect a factor of $\gtrsim{}10$ (and up to $\sim{}400$) higher  merger rate of GW150914-like systems than their formation rate. In run~D, the birth rate of new GW150914-like systems at $z\sim{}0.1$  (the redshift associated with the LIGO detection) is only 0.4  Gpc$^{-3}$ yr$^{-1}$, whereas the expected merger rate of GW150914-like systems at redshift $z\sim{}0.1$ is $\sim{}5.2$ Gpc$^{-3}$ yr$^{-1}$ (see Table~\ref{tab:table4}). This result is perfectly consistent with the estimates from the LIGO-Virgo collaboration ($3.4^{+8.8}_{-2.8}$  Gpc$^{-3}$ yr$^{-1}$, \citealt{abbott2016cO1}). 

 Fig.~\ref{fig:fig9} shows that this effect is linked to the metallicity of GW150914 progenitors. In fact, most GW150914-like systems merging at low redshift have $0.03<Z/{\rm Z}_\odot<0.05$, but the number of progenitor systems forming at low redshift in this metallicity range is a factor of $\sim{}100$ lower than the number of merging GW150914-like systems.

\begin{table}
\begin{center}
\caption{\label{tab:table4}
Expected merger rate for  GW150914, LVT151012, GW151226 and GW170104-like events. To estimate these values we consider only systems merging at redshift consistent with the LIGO detections.} 
 \leavevmode
\begin{tabular}[!h]{ccccc}
\hline
Name &  GW150914 & LVT151012 & GW151226 & GW170104\\ 
     &  [Gpc$^{-3}$ yr$^{-1}$] & [Gpc$^{-3}$ yr$^{-1}$] & [Gpc$^{-3}$ yr$^{-1}$] & [Gpc$^{-3}$ yr$^{-1}$]\\ 

\hline
D  & 5.2 & 20.9 & 8.3 & 15.9 \\
R  & 5.2 & 20.7 & 9.3 & 16.0 \\
DK  & 1.5 & 3.5 & 1.0 & 2.9 \\
D1.5  & 7.7 & 21.8 & 10.9 & 12.2 \\
D0.02  & 5.1 & 37.4 & 18.7 & 10.5\\
A16 & $3.4^{+8.8}_{-2.8}$ & $9.1^{+31}_{-8.5}$ & $36^{+95}_{-30}$ & -- \\
\noalign{\vspace{0.1cm}}
\hline
\end{tabular}
\begin{flushleft}
\footnotesize{Column 1: model name; columns 2, 3,  4, and 5: merger rate of simulated BHBs with mass and merger redshift consistent with GW150914, LVT151012, GW151226, and GW170104, respectively.  The last line shows the rate inferred from LIGO observations (see Table~II of \citealt{abbott2016cO1}) for GW150914, LVT151012, and GW151226.}
\end{flushleft}
\end{center}
\end{table}

 We find a similar trend for GW170104-like systems.  In run~D, the birth rate of new GW170104-like systems at $z\sim{}0.18$  (the redshift associated with the LIGO detection) is only 0.4  Gpc$^{-3}$ yr$^{-1}$, whereas the expected merger rate of GW170104-like systems at redshift $z\sim{}0.18$ is $\sim{}12.2$ Gpc$^{-3}$ yr$^{-1}$ (see Table~\ref{tab:table4}). 

In contrast, the shift between formation time and merger time is rather negligible for systems like GW151226. From our simulations, the expected merger rate of GW151226-like systems at redshift $z\sim{}0.1$ (the redshift associated with the LIGO detection) is $\sim{}8.3$ Gpc$^{-3}$ yr$^{-1}$ (see Table~\ref{tab:table4}), similar to  their birth rate ($\sim{}3.1$~Gpc$^{-3}$ yr$^{-1}$ at $z\sim{}0.1$). This merger rate is consistent with the estimates from the LIGO-Virgo collaboration, even if close to the low tail ($36^{+95}_{-30}$  Gpc$^{-3}$ yr$^{-1}$, \citealt{abbott2016cO1}).

 LVT151012-like events behave in a similar way to GW151226, with a mild offset between their current merger time and their current formation time (Fig.~\ref{fig:fig8}). The merger rate of LVT151012-like events is $\sim{}20$ Gpc$^{-3}$ yr$^{-1}$ at redshift $z\sim{}0.2$ (i.e. the redshift associated with the possible LIGO signal), significantly higher than that of the other two events. This happens because the predicted BHB merger rate at $z\sim{}0.2$ is higher by a factor of $\sim{}1.3$ than that at $z\sim{}0.1$ and because the mass of LVT151012 is more uncertain than that of GW151226 and GW150914 (thus, more simulated systems are consistent with it).  This merger rate is also consistent with the estimates from the LIGO-Virgo collaboration ($9.1^{+31}_{-8.5}$  Gpc$^{-3}$ yr$^{-1}$, \citealt{abbott2016cO1}).

\begin{figure}
\center{{
\epsfig{figure=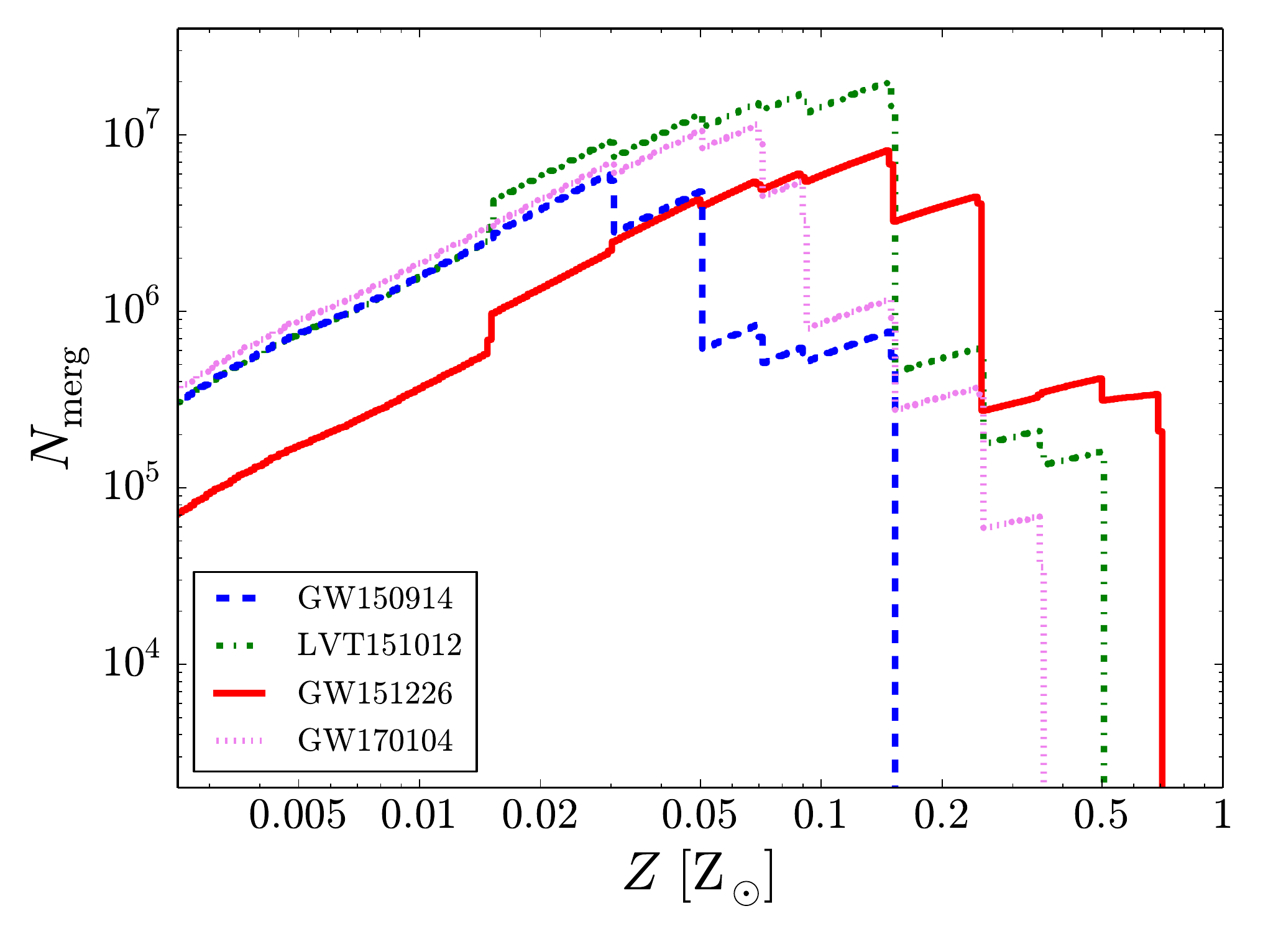,width=8cm} 
}}
\caption{\label{fig:fig10}
 Metallicity distribution of simulated GW150914-like (blues dashed line), LVT151012-like (green dash-dot line), GW151226-like  (red solid line) and GW170104-like systems (violet dotted line) in the fiducial model (run D). The value $N_{\rm merg}$ on the $y$ axis is the number of simulated BHBs per metallicity bin ($\Delta{}\log{Z}=0.01$). The step-like features in the plot correspond to the metallicity groups in the \textsc{BSE} simulations. 
}
\end{figure}

Figure~\ref{fig:fig10} shows the metallicity distribution of the stellar progenitors of   GW150914-like, GW151226-like, LVT151012-like, and GW170104-like  systems in our fiducial model. As expected, GW150914-like systems form only at relatively low metallicity: $Z\lesssim{}0.15\,{}Z_\odot$.  Metal-poor stars ($Z\lesssim{}0.3$ Z$_\odot$) are the most common progenitors also for GW170104-like systems.  In contrast, GW151226-like systems form nearly at all metallicities up to $Z\sim{}0.7$ Z$_\odot$.  LVT151012-like systems also form with a broad range of metallicities (up to $Z\sim{}0.5$ Z$_\odot$).


\section{Discussion: Comparison with previous work and caveats}\label{sec:discussion}
The method we followed in this paper ensures that we  account for the cosmic SFR and for the mass-metallicity relation, at least within the  limitations of state-of-the-art cosmological simulations. This is a crucial point for understanding the cosmic evolution of BHB mergers, since the mass of BHs is expected to depend on the metallicity of the progenitor stars.

The cosmic BHB merger rate we obtain from our models (Fig.~\ref{fig:fig1}) approximately follows the same trend as the cosmic SFR, suggesting that the slope of the curve is primarily set by star formation. The peak of the BHB merger rate is at $z\sim{}2-3$, corresponding to the peak of the cosmic SFR \citep{madaudickinson2014}. This is in reasonable agreement with previous results (e.g. \citealt{dominik2013,belczynski2016}). 

The normalization of the BHB merger rate in Fig.~\ref{fig:fig1} strongly depends on the treatment of CE (especially for HG stars) and on the distribution of  natal kicks. In particular, models in which HG donors can survive the CE phase disagree with the BHB merger rate estimated from LIGO observations, unless very large natal kicks are assumed for most BHs (in agreement with \citealt{belczynski2016}). Different distributions of the natal kicks result in a factor of $\gtrsim{}6$ different  BHB merger rate, given the large uncertainties in the distribution of BH natal kicks. Future observations by LIGO-Virgo will likely allow us to put constraints on the natal kicks of BHs \citep{vitale2016,vitale2017,stevenson2017,zevin2017}. 

The behaviour of the BHB merger rate differs significantly from the cosmic SFR only if the CE parameters are drastically different ($\alpha{}\,{}\lambda{}=0.02$) from our fiducial values ($\alpha{}\,{}\lambda{}=0.1$). 
According to the CE formalism  \citep{webbink1984}, low values of $\alpha{}\,{}\lambda{}$ imply that is more difficult to eject the envelope. 
Thus, if $\alpha{}\,{}\lambda{}$ is very low, the closest binaries merge prematurely during a CE phase, before they become BHBs. This explains why the merger rate of D0.02 is much lower than that of D at high redshift ($z>1$). 
On the other hand, a very low value of $\alpha{}\,{}\lambda{}$ implies that the shrinking of the orbit of the two cores within the CE is more efficient. Thus, looser binaries going through a CE phase might shrink enough to produce tight BHBs, which merge in a Hubble time. This might explain why the merger rate density of D0.02 increases at low redshift with respect to that of run~D: in run D0.02, there is a higher number of BHBs with long delay time (Fig.~\ref{fig:fig4}), which form at high redshift but merge at low redshift.

In contrast, if $\alpha{}\,{}\lambda{}$ is high, the CE is ejected easily and the orbit does not shrink efficiently. Thus, even if most binaries survive the CE phase, a lower number of BHBs become sufficiently close to merge in a Hubble time. These merging BHBs will have, on average, a longer delay time. This explains why less BHBs merge at high redshift ($z>0.3$) in run D1.5 with respect to run D, while  at low redshift ($z<0.3$) the merger rate of D1.5 is slightly higher than that of D.

Among previous studies,  \cite{lamberts2016} and  \cite{belczynski2016} contain several results that can be compared with ours quite straightforwardly.
 To produce their sample of BHBs, \cite{belczynski2016} use the \textsc{startrack} code \citep{belczynski2008}, while \cite{lamberts2016} use another updated version of the \textsc{BSE} code \citep{hurley2002} and focus only on the study of GW150914-like systems. The present-time BHB merger rates we obtain with $\alpha{}\,{}\lambda{}=0.1$ are consistent (within a factor of 2) with those shown by \cite{belczynski2016}, who adopt a similar choice for the CE parameters. In contrast, \cite{lamberts2016} obtain a much larger present-time BHB merger rate ($\sim{}850$ Gpc$^{-3}$ yr$^{-1}$) than expected from LIGO detections. As to the origin of this difference, we note that \cite{lamberts2016} do not update the recipes for core radii in \textsc{BSE}, and (most importantly) do not include pulsational pair instability SNe, while this paper and \cite{belczynski2016} do.



 If we restrict our analysis to systems like GW150914, GW151226, and GW170104, we find that their formation history also mimics the cosmic BHB merger rate density, but with a substantial difference. The distribution of formation times for GW150914-like and GW170104-like binaries peaks at high redshift ($z\sim{}2-4$) and then drops much faster than the total BHB merger rate density, while GW151226-like systems follow the general trend as the BHB merger rate. As a consequence, present-day GW150914-like and GW170104-like events are dominated by systems that formed at high redshift and have a long delay time. The reason is that GW150914-like and GW170104-like systems form mainly from metal-poor progenitors.  We find a rather smooth distribution of the birth times for GW150914-like systems, consistently with \cite{lamberts2016} and at odds with \cite{belczynski2016}. This is not surprising, since both us and \cite{lamberts2016} account for the mass-metallicity relation.  Our main findings for GW150914, LVT151012 and GW151226 are also consistent with the results of \cite{schneider2017}, who predict that most GW150914 candidates form at high redshift ($2.4 \leq z \leq 4.2$), whereas both GW151226 and LVT151012 have a broad range of possible formation redshifts. 

We now discuss the main caveats of the present work.  First, we have not included the impact of stellar dynamics on the evolution of BHBs. This might be a serious issue for our work, because massive stars are known to form especially in dense massive star clusters (see e.g. \citealt{weidner2006}), which are active dynamical places. Dynamical exchanges might lead to the formation of additional BHBs, which are likely more massive and more eccentric than average (e.g. \citealt{downing2010,ziosi2014,askar2016,rodriguez2016,banerjee2017}). Very massive BHs or even intermediate-mass BHs might form from runaway collisions \citep{portegieszwart2004,giersz2015,mapelli2016}. Furthermore,  Kozai-Lidov resonances in triple systems might enhance the merger rate both in dense star clusters (e.g. \citealt{kimpson2016,antonini2016}) and in the field (e.g. \citealt{antonini2017}). 
A study of the cosmic BHB merger rate including the effects of dynamics is still missing, because it poses a serious numerical challenge. 

Moreover, we neglect the contribution of very massive stars ($>150$ M$_\odot$), because the current version of \textsc{BSE} does not include them. While these very massive objects are presumably very rare, they might significantly contribute to the merger rate of the most massive BHBs (e.g. \citealt{mapelli2016}). 

Furthermore, our calculations neglect the chemically homogeneous evolutionary formation channel, recently proposed by \cite{mandel2016}, which might account for $R_{\rm BHB}\sim{}10-20$ Gpc$^{-3}$ yr$^{-1}$ (see also \citealt{marchant2016} and \citealt{demink2016}).

\begin{figure}
\center{{
\epsfig{figure=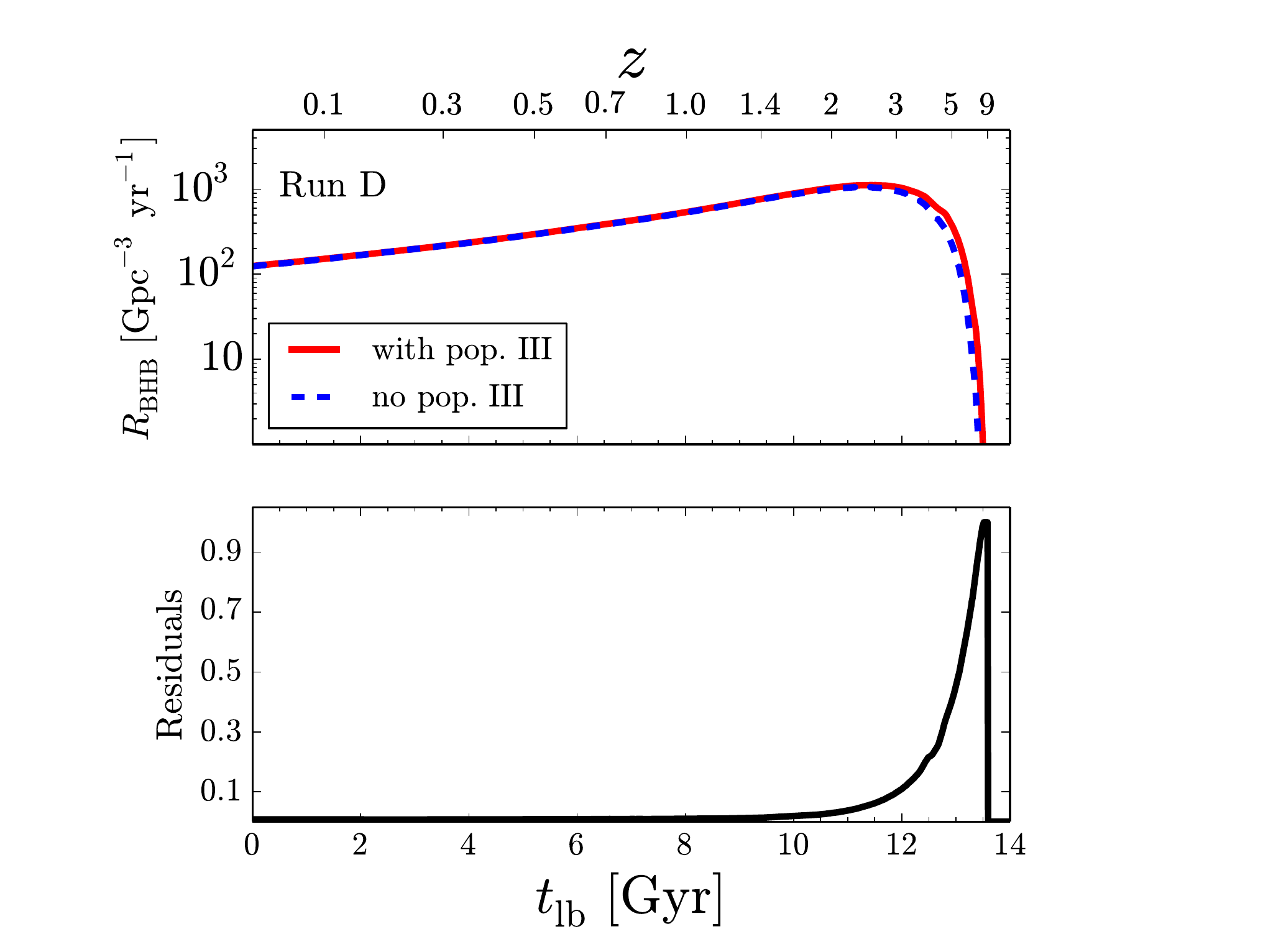,width=9cm} 
}}
\caption{\label{fig:fig11}
Upper panel: cosmic merger rate density of BHBs ($R_{\rm BHB}$) in run~D, if we include (red solid line) or neglect (blue dashed line) population~III stars (i.e. Illustris' stars with metallicity $Z\leq{}0.0001$). Lower panel: residuals of the model with population~III stars with respect to the model without them.
}
\end{figure}

In our analysis, we simply assume that population~III stars behave as ``normal'' metal-poor stars (with $Z=0.0002$). While this choice is partially motivated by the fact that stellar winds are already highly inefficient at $Z\sim{}0.0002$ (see e.g. figure 6 of \citealt{spera2015}), we do not know whether the mass function and the binary fraction of population~III stars were significantly different from those of population~II stars. On the other hand, we have checked that the contribution of population~III stars (defined  as Illustris' star particles with metallicity $Z\leq{}0.0001$) to the cosmic BHB merger rate is negligible, especially within the LIGO horizon (see Fig.~\ref{fig:fig11}). We refer to other studies (e.g. \citealt{kinugawa2014,kinugawa2016,belczynski2016b,hartwig2016,inayoshi2017}) for a more accurate treatment of BHBs from population~III stars.

We also stress that we keep using the original fitting formulas of \cite{hurley2000} for the photospheric radius and luminosity of stars, while our knowledge of the evolution of massive stars changed significantly in the last decades (e.g. \citealt{martins2013,chieffi2013,chen2015}). In forthcoming studies, we will account for more recent stellar evolution prescriptions in the frame of the new \textsc{SEVN} population-synthesis code (Spera et al., in preparation).

Finally, the Illustris reproduces  reasonably well many observational features (such as the cosmic SFR density and the galaxy luminosity function), but has several limitations which might affect the BHB merger rate. For example, it predicts a too mild decline in the cosmic SFR density at $z<1$ \citep{pillepich2017}.  This might lead to an overestimate of the present-day merger rate by several per cents (see Appendix~\ref{sec:appendix2}). Galaxies in the Illustris follow a mass-metallicity relation which is sensibly steeper than the observed relation \citep{torrey2014}. As we discuss in Appendix~\ref{sec:appendix2}, this should affect the BHB merger rate density by $\sim{}20$ per cent. Thus, it is important to repeat the same exercise that we did in this paper also with other cosmological simulations and to check for discrepancies\footnote{An updated Illustris simulation will be available soon, with an improved algorithm for AGN feedback and galactic winds \citep{pillepich2017}.}.

\section{Summary}
We reconstructed the cosmic BHB merger rate by planting BHBs into the Illustris cosmological box. The Illustris' box (length=106.5 Mpc) is large enough to guarantee that we are modelling an unbiased portion of the Universe, satisfying the cosmological principle. 

The population of BHBs is estimated through population synthesis simulations of isolated binaries, performed with an updated version of \textsc{BSE}. 
In particular, our new version of \textsc{BSE} includes up-to-date prescriptions for stellar winds of massive stars, during and after the MS. We account not only for the dependence of stellar winds on metallicity, but also for the effect of the electron-scattering Eddington limit \citep{chen2015}. Up-to-date recipes for core collapse SNe, pair-instability and pulsational pair-instability SNe are also included. 

We perform six different sets of runs with \textsc{BSE}, changing the SN prescription, the CE efficiency, the treatment of HG stars, and the distribution of natal kicks (Table~\ref{tab:table1}). Each of these six simulations produces a population of merging BHBs, depending on the metallicity of the progenitor stars. We then interface the population of merging BHBs with the Illustris simulation through a Monte Carlo model, to obtain the cosmic BHB merger rate density for each of the six \textsc{BSE} simulation sets. 

The cosmic BHB merger rate follows the same trend as the cosmic SFR, with a peak at $z\sim{}2-3$, in all simulation sets (Fig.~\ref{fig:fig1}). In contrast, the normalization of the BHB merger rate strongly depends on the specific \textsc{BSE} simulation set. In particular, the treatment of CE and the distribution of SN kicks appear to be the most important processes. 

Models in which a HG donor is allowed to survive the CE phase are not consistent with LIGO observations (unless very high natal kicks are assumed), because they give a too high BHB  present-time merger rate density. The choice of the CE parameters (we study values of $\alpha{}\,{}\lambda{}$ ranging from 0.02 to 1.5, in addition to our fiducial value $\alpha{}\,{}\lambda{}=0.1$) significantly affects the merger rate at high redshift, while it induces only minor changes (a factor of 2) in the low-redshift merger rate ($z<0.3$).

Population-synthesis simulations with large natal kicks (distributed according to \citealt{hobbs2005}) result in a present-day BHB merger rate $R_{\rm BHB}\sim{}20$ Gpc$^{-3}$ yr$^{-1}$, while population-synthesis simulations with lower kicks (accounting for the amount of fallback) give a merger rate $R_{\rm BHB}\sim{}125-155$ Gpc$^{-3}$ yr$^{-1}$. Both rates are still consistent with the constraints from LIGO observations ($9-240$  Gpc yr$^{-1}$, \citealt{abbott2016cO1}), but forthcoming detections might be able to distinguish between them.

From our simulations, we can also trace the metallicity of the progenitors of merging BHBs. 
We find that most BHB mergers detectable by current GW interferometers 
come from relatively metal-poor progenitors, ranging from $\sim{}0.015$ Z$_\odot$ to $\sim{}0.2$ Z$_\odot$ (Figs~\ref{fig:fig2} and \ref{fig:fig7}).

The merging BHBs have chirp masses (total masses) ranging from $\sim{}3$ to $\sim{}35$ M$_\odot$ ($\sim{}6$ to $\sim{}82$ M$_\odot$), with a slight dependence on the SN prescription and on CE parameters (Fig.~\ref{fig:fig6}). 

If we focus on systems similar  to GW150914, GW151226 and GW170104, we can obtain some useful hints on their progenitors. The formation of GW150914-like and GW170104-like systems is more efficient at high redshift and drops in the local Universe (Fig.~\ref{fig:fig8}). Most GW150914-like and GW170104-like systems merging in the local Universe appear to have formed at higher redshift with a long delay time.  This happens because only genuinely metal-poor stars  can produce GW150914-like and GW170104-like systems (with metallicity $Z<0.15$ Z$_\odot$ and $Z<0.3$ Z$_\odot$, respectively; see Fig.~\ref{fig:fig10}). In contrast, GW151226-like systems form and merge all the way through the cosmic history (from $z\sim{}9$ to $z=0$). The progenitors of GW151226-like systems can be either metal-rich or metal-poor stars with about the same probability.

In our fiducial model (run~D) the percentage of GW150914-like systems merging within the LIGO instrumental horizon is higher ($\sim{}6$ \%) than the percentage of GW151226-like systems ($\sim{}3$ \%, see Table~\ref{tab:table3}). This might suggest that LIGO and Virgo will observe more events like GW150914 than like GW151226.

In conclusion, this  study provides several clues about merging BHBs and their progenitors. We show that the BHB merger rate density poses constraints on both the CE process and the natal kicks of BHs. Forthcoming GW detections will allow us to further strengthen these constraints. Moreover, our results suggest that most BHBs merging within the LIGO horizon formed from relatively metal-poor progenitors ($Z<0.2$ Z$_\odot$). The Illustris simulation also includes information on the properties of BHB host galaxies. In a forthcoming paper, we will explore the properties of the galaxies  where BHBs form and merge. The same analysis will be done also for NS binary systems and NS-BH binary systems, providing a clue for the detection of electromagnetic counterparts.


\section*{Acknowledgments}
We thank the anonymous referee for their critical reading of the manuscript. We thank Alessandro Bressan, Marica Branchesi, Raffaella Schneider, Luca Graziani, Elena D'Onghia and Roberto Maiolino for useful discussions. We warmly thank The Illustris team for making their simulations publicly available. Numerical calculations have been performed through a CINECA-INFN agreement and through a CINECA-INAF agreement, providing access to resources on GALILEO and MARCONI at CINECA.  
MM and MS acknowledge financial support from the Italian Ministry of Education, University and Research (MIUR) through grant FIRB 2012 RBFR12PM1F, and from INAF through grant PRIN-2014-14. MM acknowledges financial support from the MERAC Foundation.

\bibliography{./bibliography}
\appendix
\section{Prescriptions for core-collapse SN and for fallback}\label{sec:appendix1}
 We adopt two different models for core-collapse SNe: the rapid (R) and the delayed (D) model. Both models were introduced by \cite{fryer2012} and they differ by the time-scale over which the explosion occurs: $<250$ ms after the bounce for the rapid model, $> 250$ ms for the delayed mechanism.

\subsubsection{Rapid SN model}

For the rapid SN mechanism, a fixed mass of the proto-compact object, $m_{\rm proto}=1.0\,{}{\rm M}_{\odot}$, is assumed.
In this case, the fallback parameter $f_{\rm fb}$ is defined as

\begin{equation}
f_{\rm fb}=
\begin{cases}
0.2\,{}(m_{\rm fin}-m_{\rm proto})^{-1} &  m_{\rm CO}<2.5\,{}{\rm M}_{\odot} \\
\dfrac{0.286\,{}m_{\rm CO}-0.514}{m_{\rm fin}-m_{\rm proto}} &  2.5\,{}{\rm M}_{\odot}\leq{}m_{\rm CO} <6.0\,{}{\rm M}_{\odot} \\
1.0 & 6.0\,{}{\rm M}_{\odot} \leq{} m_{\rm CO} < 7.0\,{}{\rm M}_{\odot} \\
\alpha_{\rm R}\,{}m_{\rm CO}+ \beta_{\rm R} &  7.0\,{}{\rm M}_{\odot}\leq{} m_{\rm CO} <11.0\,{}{\rm M}_{\odot} \\
1.0 & m_{\rm CO} \geq 11.0\,{}{\rm M}_{\odot}
\label{eq:rapid}
\end{cases}
\end{equation}where $m_{\rm fin}$ is the final mass of the star, $m_{\rm proto}$ is the mass of the proto-NS, $m_{\rm CO}$ is the mass of the Carbon-Oxygen core, $\alpha_{\rm R}\equiv 0.25 - 1.275\,{}(m_{\rm fin}-m_{\rm proto})^{-1}$, and $\beta_{\rm R}\equiv 1-11\alpha_{\rm R}$.

For all models, the  amount of mass that falls onto the proto-NS is defined as  $m_{\rm fb}=f_{\rm fb}\,{}\left(m_{\rm fin}-m_{\rm proto}\right)$.  The final mass of the compact object is given by $M_{\textrm{BH}}=m_{\rm proto}+m_{\rm fb}$.

\subsubsection{Delayed SN model}
For the delayed SN mechanism, the mass of the proto-compact object is defined as

\begin{equation}
m_{\rm proto}=
\begin{cases}
1.2 \,{}{\rm M}_{\odot} &  m_{\rm CO}<3.5\,{}{\rm M}_{\odot} \\
1.3 \,{}{\rm M}_{\odot} &  3.5\,{}{\rm M}_{\odot}\leq{}m_{\rm CO} <6.0\,{}{\rm M}_{\odot} \\
1.4 \,{}{\rm M}_{\odot} &  6.0\,{}{\rm M}_{\odot}\leq{}m_{\rm CO} <11.0\,{}{\rm M}_{\odot} \\
1.6 \,{}{\rm M}_{\odot} &  m_{\rm CO} \geq 11.0 \,{}{\rm M}_{\odot}.
\end{cases}
\end{equation}
The amount of fallback is determined using the following relations

\begin{equation}
f_{\rm fb}=
\begin{cases}
\dfrac{0.2}{m_{\rm fin}-m_{\rm proto}} &  m_{\rm CO}<2.5\,{}{\rm M}_{\odot} \\
\dfrac{0.5\,{}m_{\rm CO}-1.05\,{}{\rm M}_{\odot}}{m_{\rm fin}-m_{\rm proto}} &  2.5\,{}{\rm M}_{\odot}\leq{}m_{\rm CO} <3.5\,{}{\rm M}_{\odot} \\
\alpha_{\rm D}\,{}m_{\rm CO}+ \beta_{\rm D} &  3.5\,{}{\rm M}_{\odot}\leq{}m_{\rm CO} <11.0\,{}{\rm M}_{\odot} \\
1.0 & m_{\rm CO} \geq 11.0\,{}{\rm M}_{\odot}
\label{eq:delayed}
\end{cases}
\end{equation}
where $\alpha_{\rm D}\equiv 0.133 - 0.093\,{}(m_{\rm fin}-m_{\rm proto})^{-1}$ and $\beta_{\rm D}\equiv 1-11\alpha_{\rm D}$.

\section{The impact of resolution and mass-metallicity relation}\label{sec:appendix2}
\begin{figure}
\center{{
\epsfig{figure=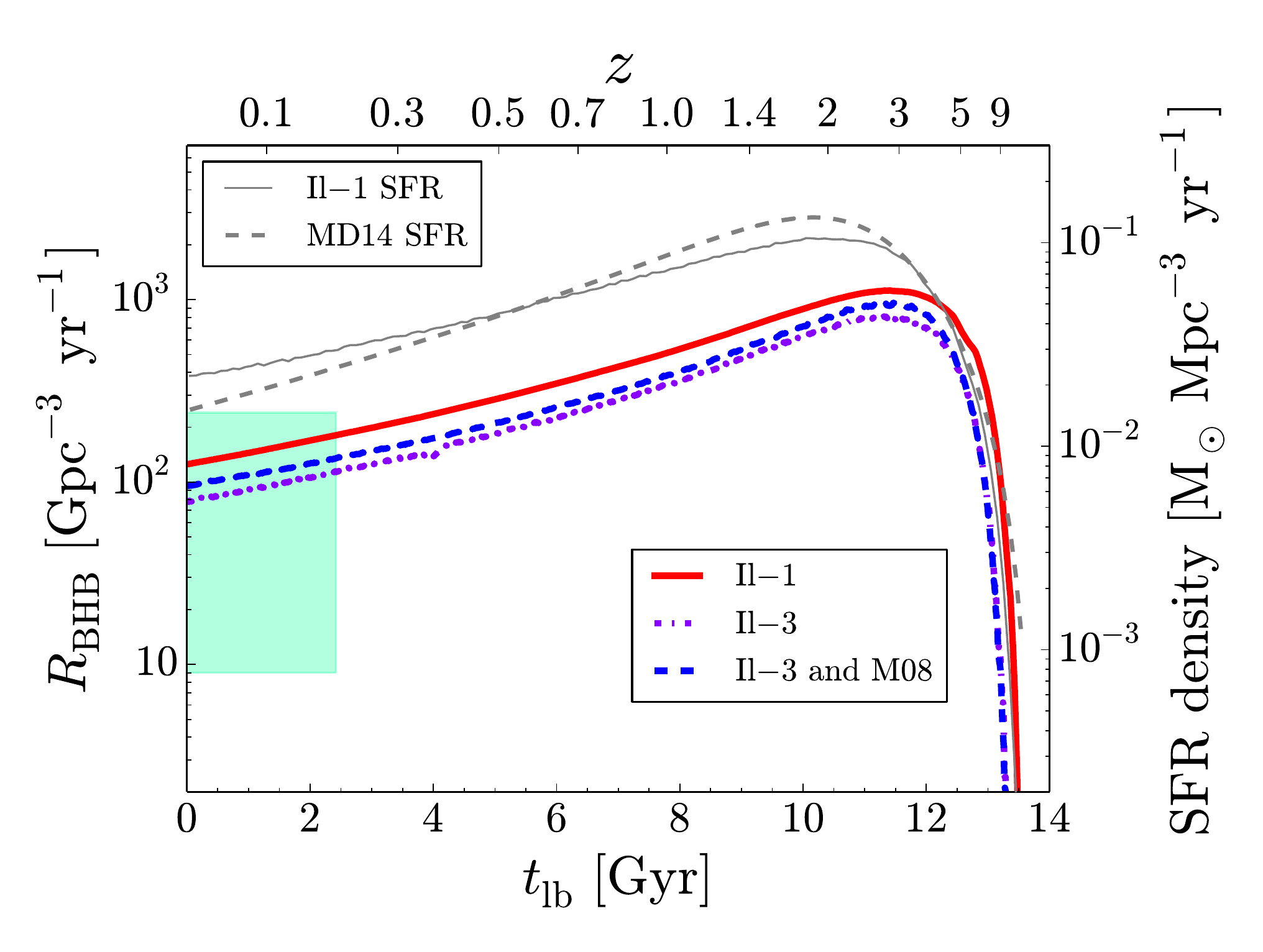,width=9cm} 
}}
\caption{\label{fig:figB1}
Left $y-$axis: cosmic merger rate density of BHBs ($R_{\rm BHB}$) in the comoving frame, as a function of the look-back time $t_{\rm lb}$ (bottom $x$ axis) and of the redshift $z$ (top $x$ axis). Red solid line: model D applied to the Illustris-1 simulation (same as in Fig.~\ref{fig:fig1}); violet dash-dot line: model D applied to the Illustris-3 simulation; blue dashed line: model D applied to the Illustris-3 simulation and adopting Maiolino et al. (2008; hereafter M08) fitting formulas for the mass-metallicity relation. Green shaded area: BHB merger rate inferred from LIGO detections \citep{abbott2016cO1}. Right $y-$axis: cosmic SFR density from the Illustris-1 (grey thin solid line) and from \citealt{madaudickinson2014} (grey thin dashed line), as a function of the look-back time $t_{\rm lb}$ (bottom $x$ axis) and of the redshift $z$ (top $x$ axis). 
}
\end{figure}

\begin{table}
\begin{center}
\caption{\label{tab:tableB1}
Comoving BHB merger-rate density at redshift $z=0$ and $z=0.2$.} 
 \leavevmode
\begin{tabular}[!h]{ccc}
\hline
Name &  $R_{\rm BHB} (z=0)$ & $R_{\rm BHB}(z=0.2)$\\ 
     & [Gpc$^{-3}$ yr$^{-1}$] & [Gpc$^{-3}$ yr$^{-1}$]\\
\hline
Il-1  & 125 & 181\\
Il-3  & 78 & 114 \\
Il-3 and M08 & 96 & 135 \\
\noalign{\vspace{0.1cm}}
\hline
\end{tabular}
\begin{flushleft}
\footnotesize{Same as Table~\ref{tab:table1} but for the check runs. Column 1: model name; column 2: present-time BHB merger rate density; column 3: BHB merger rate density at $z=0.2$.}
\end{flushleft}
\end{center}
\end{table}

\begin{table}
\begin{center}
\caption{\label{tab:tableB2}
Best fit parameters for the mass-metallicity relation in equation~\ref{eq:maiolino} at different redshift. The values of $M_0$ and $K_0$ at $z=3.5$ come from Mannucci et al. (2009), while the other values come from Maiolino et al. (2008).}
 \leavevmode
\begin{tabular}[!h]{ccc}
\hline
$z$ &  $\log{M_0}$ & $K_0$\\ 
\hline
0.07  & 11.18 & 9.04 \\
0.7   & 11.57 & 9.04 \\
2.2   & 12.38 & 8.99 \\
3.5  &  12.28  & 8.69    \\
\noalign{\vspace{0.1cm}}
\hline
\end{tabular}
\begin{flushleft}
\footnotesize{Column~1: redshift; column~2 and 3: values of the parameters in equation~\ref{eq:maiolino} at different redshift.}
\end{flushleft}
\end{center}
\end{table}

In this section, we estimate the impact of the numerical resolution and of the simulated mass-metallicity relation on the main results of this paper. In Fig.~\ref{fig:figB1} and Table~\ref{tab:tableB1}, we compare the BHB merger rate density we obtain from the Illustris-1 simulation (model Il-1) with the BHB merger rate density we obtain from the Illustris-3 simulation (model Il-3), which has a factor of $\sim{}60$ poorer resolution. The current BHB merger rate density in the Illustris-3 simulation is $\sim{}40$~\% lower than in the Illustris-1 simulation, because of the lower resolution. Galaxies with stellar mass $<10^9$ M$_\odot$ are under-resolved (they consist of $<1000$ star particles) even in the Illustris-1 simulation. It is reasonable to expect that accounting for these low-mass galaxies might further boost the merger rate.

In a companion paper \citep{schneider2017}, we follow a complementary approach by coupling our population-synthesis models with the \textsc{GAMESH} pipeline \citep{graziani2015,graziani2017}. \textsc{GAMESH} consists of a  $(4\,{}{\rm Mpc})^3$ box and reaches a much higher resolution: even galaxies with a stellar mass of $\sim{}10^6$ M$_\odot$ are effectively resolved. \cite{schneider2017} predict that all GW150914-like systems in the \textsc{GAMESH} box form in progenitor galaxies with a stellar mass $<5\times{}10^6$ M$_\odot$ (see their Figure~2). Thus, the merger rate of GW150914-like systems is probably underestimated in the current study, due to the Illustris resolution.


Fig.~\ref{fig:figB1} also shows the difference between the cosmic SFR density in the Illustris-1 simulation and the cosmic SFR density from \cite{madaudickinson2014}. At redshift $z<0.3$ the Illustris-1 simulation predicts a higher SFR density by a factor of $\sim{}1.5$. This suggests that the current BHB merger rate density might be slightly overestimated in our calculations.

As we discussed in Section~\ref{sec:section2.2}, the model of sub-grid physics adopted in the Illustris produces a mass-metallicity relation  which significantly differs from the observed one. In particular, the simulated mass-metallicity relation  is sensibly steeper than the observed one and does not show the observed turnover at high stellar mass \citep{vogelsberger2013,torrey2014}. We quantify the impact of these differences between simulated and  observed mass-metallicity relation through the following procedure.

In the Illustris-3 simulation, we override the metallicity of a given star particle with the metallicity we expect from the observed mass-metallicity relation. For the observed relation, we adopt the fitting formula by \cite{maiolino2008} and \cite{mannucci2009}:
\begin{equation}\label{eq:maiolino}
12+\log{[{\rm O}/{\rm H}]}=-0.0864\,{}(\log{M_\ast{}} - \log{M_0})^2\,{}+K_0,
\end{equation} 
where $M_\ast$ is the total stellar mass of the host galaxy in solar masses, while $M_0$ and $K_0$ are given in Table~\ref{tab:tableB2}. For intermediate redshifts between those in Table~\ref{tab:tableB2}, we obtain the metallicity by linear interpolation. At redshift $z<0.07$ ($z>3.5$) we simply use the same coefficients as for $z=0.07$ ($z=3.5$). The new metallicity of each Illustris' star is randomly extracted from a Gaussian distribution with mean value given by equation~\ref{eq:maiolino} (where $M_\ast{}$ is the total stellar mass of the sub-halo hosting the Illustris' star) and standard deviation $\sigma{}=0.5$ dex (accounting for metallicity dispersion within galaxies). 

In Fig.~\ref{fig:figB1} and Table~\ref{tab:tableB1} we show the BHB merger rate density we obtain with this procedure, compared to the BHB merger rate density we derive using the simulated metallicity of each Illustris-3 particle. The maximum difference between the two curves is only $\sim{}20$ per cent. The same procedure can be applied to the Illustris-1 simulation, but is a factor of 60 more computationally expensive. We will show the results for the Illustris-1 in our follow-up paper.


\end{document}